\documentclass[fleqn,usenatbib]{mnras}

\usepackage{newtxtext,newtxmath}
\usepackage[T1]{fontenc}

\DeclareRobustCommand{\VAN}[3]{#2}
\let\VANthebibliography\thebibliography
\def\thebibliography{\DeclareRobustCommand{\VAN}[3]{##3}\VANthebibliography}

\usepackage{graphicx}	% Including figure files
\usepackage{amsmath}	% Advanced maths commands
\usepackage{multirow}
\usepackage{multicol}
\usepackage{booktabs}

\usepackage[dvipsnames]{xcolor} % for text colours
\newcommand{\NuSTAR}[0]{\textit{NuSTAR}}
\newcommand{\XMM}[0]{\textit{XMM-Newton}}
\newcommand{\RXTE}[0]{\textit{RXTE}}
\newcommand{\fekalpha}{Fe~K$\alpha$}
\newcommand\rg{\text{r}_\text{g}}
\newcommand\Msun{\text{M}_\odot}

\defcitealias{Ingram2022}{I22}
\defcitealias{Steiner2012}{S12}
\title[H1743-322 mass measurement]{Proof of principle X-ray reflection mass measurement of the black hole in H1743-322}

\author[E. Nathan et al.]{Edward Nathan$^{1}$, 
Adam Ingram$^{2}$, 
James F. Steiner$^{3}$,
Ole K\"onig$^{3,4}$, 
Thomas Dauser$^4$, 
Matteo Lucchini$^5$,
\newauthor
Guglielmo Mastroserio$^6$, 
Michiel van der Klis$^5$,
Javier A. Garc\'ia$^{7,1}$, 
Riley Connors$^8$, 
Erin Kara$^9$,
Jingyi Wang$^9$
\\
$^{1}$California Institute of Technology, Pasadena, CA 91125, USA\\
$^{2}$School of Mathematics, Statistics, and Physics, Newcastle University, Newcastle upon Tyne NE1 7RU, UK\\
$^{3}$Center for Astrophysics \textbar\ Harvard \& Smithsonian, 60 Garden St, Cambridge, MA 02138, USA\\
$^4$Dr. Karl Remeis-Observatory and Erlangen Centre for Astroparticle Physics, Sternwartstr. 7, 96049 Bamberg, Germany\\
$^5$Anton Pannekoek Institute, University of Amsterdam, Science Park 904, 1098 XH Amsterdam, The Netherlands\\
$^6$Dipartimento di Fisica, Universit\`{a} degli Studi di Milano, Via Celoria 16, I-20133 Milano, Italy\\
$^7$X-ray Astrophysics Laboratory, NASA Goddard Space Flight Center, Greenbelt, MD 20771, USA\\
$^8$Villanova University, Department of Physics, Villanova, PA 19085, USA \\
$^9$MIT Kavli Institute for Astrophysics and Space Research, MIT, 70 Vassar Street, Cambridge, MA 02139, USA
}

\date{Accepted XXX. Received YYY; in original form ZZZ}

\pubyear{2024}

\begin{document}
\label{firstpage}
\pagerange{\pageref{firstpage}--\pageref{lastpage}}
\maketitle

% Abstract of the paper
\begin{abstract}
The black hole X-ray binary H1743-322 lies in a region of the Galaxy with high extinction, and therefore it has not been possible to make a dynamical mass measurement.  In this paper we make use of a recent model which uses the X-ray reflection spectrum to constrain the ratio of the black hole mass to the source distance. By folding in a reported distance measurement, we are able to estimate the mass of the black hole to be $12\pm2~\Msun{}$ ($1\sigma$ credible interval).  
We are then able to revise a previous disc continuum fitting estimate of black hole spin $a_*$ (previously relying on a population mass distribution) using our new mass constraint, finding $a_*=0.47\pm0.10$.
This work is a proof of principle demonstration of the method, showing it can be used to find the mass of black holes in X-ray binaries.
\end{abstract}

% Select between one and six entries from the list of approved keywords.
% Don't make up new ones.

%%%%%%%%%%%%%%%%%%%%%%%%%%%%%%%%%%%%%%%%%%%%%%%%%%

%%%%%%%%%%%%%%%%% BODY OF PAPER %%%%%%%%%%%%%%%%%%

\section{Introduction}

Accreting black holes (BHs) have been seen with masses spanning many orders of magnitude, from stellar mass BHs in X-ray binaries (XRBs) which can be as small as a few solar masses, all the way to supermassive BHs in active galactic nuclei (AGN) which can have masses up to $\sim 10^{10}~\Msun{}$.  

XRB systems, which consist of a stellar-mass BH and a stellar companion, glow brightly in X-rays during epochs of increased mass transfer onto the BH. The X-ray spectrum we see contains a thermal accretion disc \citep{Shakura1973,Novikov1973} alongside a hard X-ray power law, thought to be generated by Compton up-scattering of disc photons in a cloud of hot electrons typically referred to as the \textit{corona} \citep{Thorne1975, Sunyaev1979}.  However, the origin and the geometry of the corona is still an active area of research \citep[see e.g.][]{Done2007,Poutanen2018,Bambi2021,Kammoun2024}.
In addition to thermal disc and power law coronal emission, we see \textit{reflection} features associated with high-energy coronal photons which are incident upon the disc where they are reprocessed and scattered back into our line of sight.  These reprocessed photons form a reflection spectrum that includes many atomic features such as the prominent Fe~K complex (including the strong \fekalpha{} emission line) at $\sim6.4-6.9$~keV, the Fe~K absorption edge at $\sim 7.1$~keV, and a Compton hump at $\sim20-30$~keV from electron scattering \citep{Matt1991,Ross2005,Garcia2013}.  These features are created in the local rest-frame of the disc, however as an observer we see the disc in a highly relativistic regime; thus crucially we see different patches of the disc with different Doppler and gravitational blue/red-shifts. Therefore, the overall reflection spectrum we see is broadened and shifted in a way that traces the accretion flow \citep{Fabian1989,Dauser2013}.

Most XRBs detected in the Milky Way are transient systems which spend most of their time in quiescence during which they are dim and difficult to detect, with occasional episodes of enhanced mass transfer causing bright outbursts.  During these outbursts the X-ray spectrum evolves: beginning in the hard state, dominated by the hard coronal power law, with the flux rising over a timescale of days to months.  Eventually the source transitions to the disc-dominated soft state, via the intermediate state \citep{Fender2004, Remillard2006,Done2007, Belloni2010, Fender2012}.

When an XRB is in quiescence, the BH mass can be measured dynamically by tracking the orbital phase-dependent Doppler shifts of absorption lines in the optical spectrum of the companion star \citep[e.g.][]{Casares2014}.  The BH mass has been measured in this way for $\sim20$ out of a total of $\sim70$ known BH XRBs \citep{Corral-Santana2016}.  These optical techniques can be inhibited by extinction, an issue that primarily effects sources within the Galactic plane and bulge.  
\citet{Jonker2021} hypothesised that this could cause a bias to the measured BH mass population: assuming XRBs are likely to form in the Galactic plane or bulge, lower mass systems may be more likely to receive a large enough `kick' during the BH formation to push the system away from the regions with higher extinction in which they were originally formed \citep{Fryer2001}. Therefore, this suggests that we are better able to measure the BH mass in lower mass systems, generating the bias.  
\citet{Jonker2021} further suggested that this could be a reason why the population of BHs detected through gravitational waves (during mergers) is more massive than those detected in XRBs \citep{Farr2011,Corral-Santana2016,Abbott2021}, even after correcting for the selection effect of gravitational wave detectors being more sensitive to higher mass BHs \citep[although it may be that the population of BHs in low mass XRB are consistent with the binary black hole population for which the secondary BH mass is $<8~\Msun$, which could be a better comparison;][]{Fishbach2022}. 

BH mass measurements using the X-ray signal are not susceptible to the potential extinction bias, and therefore could in principle provide a test of the \citet{Jonker2021} hypothesis. 
Whereas the energy shifts to the iron line are only sensitive to size scales in units of the gravitational radius $\rg=GM/c^2$ and are therefore insensitive to mass, other properties of the X-ray signal can be used to infer the mass. Joint modelling of the reflection spectrum and the reverberation delay between fluctuations in the direct and reflected energy signals yields a BH mass measurement \citep{Mastroserio2019}. \citet[][hereafter I22]{Ingram2022} proposed that such an analysis could additionally be used to measure the distance $D$ to the system from Earth. This is because the shape of the restframe reflection spectrum depends on the total reflected flux that is emitted by the accretion disc \citep{Garcia2016,Tomsick2018, Jiang2019a, Jiang2019b, Garcia2019}. 
Reverberation mapping constrains the emitting area of the disc by introducing physical units from the light-crossing delays; therefore the intrinsic reflected luminosity is known and the distance can be inferred from the observed reflected flux, breaking the degeneracy between $M$ and $D$.
In the case that only the X-ray spectrum is considered, then the emitting area of the disc is only known in units of $\rg^2 \propto M^2$. Therefore reflection spectroscopy alone can be used to constrain the ratio $M/D$. Thus, an existing mass measurement can be used in combination with the X-ray spectrum to infer the distance, or conversely an existing distance measurement can be utilized to infer the BH mass.

H1743-322 (hereafter H1743) is a well-studied BH XRB which has undergone many outbursts \citep{Tetarenko2016}.  However, it is in a region of high extinction close to the Galactic centre, and therefore no optical dynamical mass measurement has been possible. 
There have been attempts to use the observed X-rays to constrain the BH mass before:  
\citet{Petri2008} used high frequency quasi-periodic oscillations (QPOs) to estimate the BH mass in H1743 to be between $9$ and $13~\Msun{}$.  
\citet{Shaposhnikov2009} used the frequency of low frequency QPOs, and spectral properties, to estimate the BH mass as $13.3\pm3.2~\Msun{}$; using the known mass, distance, and inclination of the XRB GRO~J1655-40 as a reference.  
\citet{Ingram2014} used an observed doublet of high frequency QPOs,  along with the highest observed frequency of a Type-C low frequency QPO to place a lower bound on the BH mass $\geq9.29~\Msun{}$ with the relativistic precession model.  
\citet{Molla2017} used the frequency of low frequency QPOs, also in comparison to GRO~J1655-40, along with spectral fits using the Two-Component Advective Flow (\textsc{tcaf}) model, to report a mass estimate of $11\pm2~\Msun{}$.  Finally, \citet{Tursunov2018} have also used QPO analysis to suggest the mass lies between $7.4\text{ and }13.6~\Msun{}$.

The distance to H1743 was measured by \citet[][hereafter S12]{Steiner2012} by tracking the motion of approaching and receding super-luminal jet ejecta with radio observations.  They reported the system was at $D=8.5\pm0.8$~kpc, consistent with the likely scenario of the source residing in the Galactic bulge.
This distance measurement means it is therefore possible to obtain a mass measurement of this source with reflection spectroscopy. In this paper, we fit the spectrum with \textsc{rtdist}, the newest version of the \textsc{reltrans} model \citep{Ingram2019reltrans} that is described in \citetalias{Ingram2022}, and input the \citetalias{Steiner2012} distance measurement as a Bayesian prior to constrain the BH mass. We utilise long-look \XMM{} and \NuSTAR{} observations of the source from five epochs while it was in the hard state during its 2014, 2016, and 2018 outbursts. This constitutes the first test of the new \textsc{reltrans} model, and whether it can reliably measure the masses of BHs in extincted XRBs using the X-ray reflection spectrum. Moreover, we use our mass measurement to revise the BH spin measurement originally obtained by \citetalias{Steiner2012} by modelling the thermal accretion disc in the soft state. The X-ray continuum-fitting approach for measuring spin requires an external estimate of distance and mass. Whereas \citetalias{Steiner2012} used the mass distribution known at the time for the \textit{population} of XRBs, we now instead use our mass measurement for this particular source.

The outline of the paper is as follows: in Section~\ref{sec:observations} we describe the observations used and our data analysis procedures, and present a summary of the X-ray timing properties; Section~\ref{sec:fits} describes our spectral fitting procedure and presents our results, including the mass posterior; and Section~\ref{sec:spin} presents our revised spin measurement and we respectively discuss our results and draw conclusions in Sections~\ref{sec:discussion} and~\ref{sec:conclusions}.

\section{Observations and Data Reduction}
\label{sec:observations}
\subsection{Observation Log}
We use observations of three outbursts of H1743 which occurred in 2014, 2016, and 2018. We analyze five observations from the \textit{Nuclear Spectroscopic Telescope ARray} \citep[\NuSTAR{};][]{Harrison2013}, along with five observations from \textit{X-ray Multi-Mirror Mission} \citep[\XMM{};][]{Jansen2001}. 
We consider these as five epochs, which we label 14, 16a, 16b, 18a, and 18b, based upon the year of the \NuSTAR{} observation. The details of the ten different observations used are listed in Table~\ref{tab:observations}.  The \XMM{} and \NuSTAR{} data from epoch 14 were also included in \citet[][named `XMM2a' and `XMM2b']{Ingram2016, Ingram2017h1743}.

\begin{table*}
    \centering
    \caption{Details of the observations used, with the start and end time formats given as "YY-MM-DD hh:mm" in UTC.  The observations are split into five epochs of simultaneous observations, each with one \NuSTAR{} and between zero to two \XMM{} observations, as listed.}
    \begin{tabular}{rcccccc}
\toprule
 \multirow{2}{*}{Epoch} & \multicolumn{3}{c}{\textit{NuSTAR}}  & \multicolumn{3}{c}{ \textit{XMM-Newton}}  \\
 & ObsID & Start  & End & ObsID & Start & End \\
\midrule
 \multirow{2}{*}{14}  & \multirow{2}{*}{80001044004} & \multirow{2}{*}{14-09-23 18:21} & \multirow{2}{*}{14-09-25 08:51}  & 0724401901 & 14-09-23 20:08 & 14-09-24 18:00\\
                      &                              &                                 &                                  & 0740980201 & 14-09-24 18:54 & 14-09-25 08:35\\
                  16a &                  80202012002 &                  16-03-13 01:36 &                   16-03-14 19:11 & 0783540201 & 16-03-13 04:18 & 16-03-14 19:00\\
                  16b &                  80202012004 &                  16-03-15 06:46 &                   16-03-17 00:21 & 0783540301 & 16-03-15 05:00 & 16-03-16 18:52\\
                  18a &                  90401335002 &                  18-09-19 02:36 &                   18-09-20 01:01 & \multicolumn{3}{c}{---} \\ 
                  18b &                  80202012006 &                  18-09-26 08:36 &                   18-09-27 21:31 & 0783540401 & 18-09-26 14:39 & 18-09-28 03:04\\ 
\bottomrule
\end{tabular}
    \label{tab:observations}
\end{table*}

\subsection{\NuSTAR{} Data Reduction}

We used the \NuSTAR{} data reduction software \textsc{NuSTARDAS v2.0.0}, with \textsc{CALDB v20200726}.  We first used the \textsc{FTOOL} \textsc{nupipline} to produce cleaned event lists, before extracting source and background products with the \textsc{FTOOL} \textsc{nuproducts}.  For each observation, we used a $60"$ circular source extraction region centered on the bright source, and a $90"$ circular background extraction region chosen from an area of the detector without source counts.  Finally, we used the \textsc{grppha} to bin the energy spectra so that they contain a minimum of $25$ counts per channel.

\subsection{\XMM{} Data Reduction}
The \XMM{} observations were taken with the Reflection Grating Spectrometers (RGS) and the European Photon Imaging Camera (EPIC-pn) in timing mode to avoid pile-up due to the brightness of the source. There was no data recorded with the MOS cameras. We used the \XMM{} Science Analysis Software (SAS) version 20.0.0.

\subsubsection{EPIC-PN Timing Mode}
\label{sec:XMM_EPIC_pn_reduction}
We used the SAS task \textsc{epproc} to extract cleaned event lists, before extracting source products from a region defined with $31\leq\text{RAWX}\leq45$, using only single and double events ($\text{PATTERN} \leq 4$), and ignoring bad pixels ($\text{FLAG}=0$). We rebinned the spectra to have a minimum of 25 counts per channel using the SAS tool \textsc{specgroup}. We generated the response using \textsc{rmfgen} using the calibration published in the current calibration file `\textsc{XRT3\_XAREAEF\_0014.CCF}', which is available with the setting \textsc{applyabsfluxcorr=yes}. The ancillary response file was generated with \textsc{arfgen}.

\subsubsection{RGS}
We used the SAS task \textsc{rgsproc} to produce the source and background spectra, along with the response, for the first spectral order of each RGS1 and RGS2.  The RGS spectra were also rebinned using \textsc{specgroup} to a minimum of 25 counts per channel.

\subsection{Cross-Calibration}

\begin{table}
    \centering
    \caption{The difference in the power law index and model normalisations needed to cross-calibrate the \XMM{} observations with their simultaneous \NuSTAR{} counterparts, when a simple absorbed power law model is fit to the overlapping $3-10$~keV band. The values are given are in comparison with the \NuSTAR{} FMPA values, such that the \XMM{} data is steeper with a lower normalisation.  The representative parameter uncertainties are $3\times10^{-3}$ for both $\Delta\Gamma$ and the relative normalisations.}
    \begin{tabular}{rcccc}
\toprule
                  Epoch &              \textit{NuSTAR} & \textit{XMM-Newton} & $\Delta\Gamma$ & Rel. norm. \\
\midrule
    \multirow{2}{*}{14} & \multirow{2}{*}{80001044004} &          0724401901 &          0.106 &      0.756 \\
                        &                              &          0740980201 &          0.111 &      0.756 \\
                    16a &                  80202012002 &          0783540201 &          0.088 &      0.777 \\
                    16b &                  80202012004 &          0783540301 &          0.094 &      0.768 \\
                    18b &                  80202012006 &          0783540401 &          0.085 &      0.751 \\ 
\bottomrule
\end{tabular}
    \label{tab:cross_calib}
\end{table}

\begin{figure}
    \centering
    \includegraphics[width=\columnwidth,trim={0mm 30mm 0mm 20mm}]{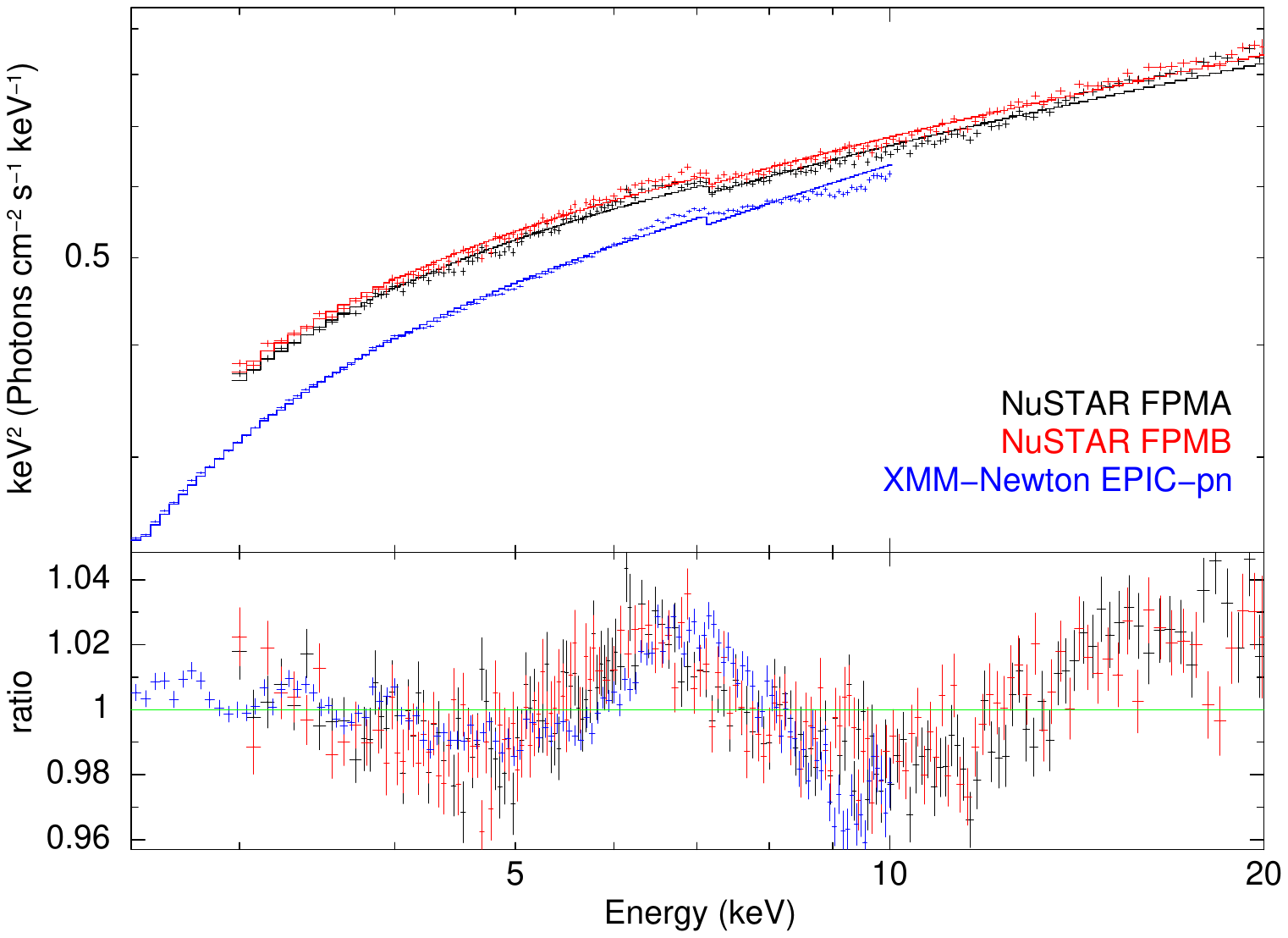}
    \caption{The $3-20$~keV flux-energy spectra from \NuSTAR{}'s FMPA (black) and FMPB (red) detectors, along with \XMM{}'s EPIC-pn (blue) which was in timing mode, from the 16a epoch.  These were simultaneously fit to an absorbed power law, each with a free normalisation, and with the power law index tied between FMPA and FMPB, however with the EPIC-pn index otherwise free.  The top panel shows the unfolded data and model, while the lower panel shows the ratio of the data to the absorbed power law model. This shows the strong Fe relativistic reflection feature at $\sim6-8$~keV}.
    \label{fig:X-calibration}
\end{figure}

The observations we consider come from five separate instruments,mand therefore we make use of different energy ranges. Throughout this paper, using \NuSTAR{}'s FPMA and FPMB, we cover 3--78~keV; with \XMM{}, using EPIC-pn (in timing mode) we cover 2.5--10~keV, and using RGS1 and RGS2 we cover 0.9--2.1~keV.

As the energy range of \XMM{}'s EPIC-pn camera and \NuSTAR{}'s FPMs overlap, it is important that we check the cross-calibration between them.  
To do this, we fit the spectra from the three instruments with a simple absorbed power law model, considering only the overlapping energy range ($3-10$ keV). For line-of-sight absorption, we use \textsc{tbabs} with the abundances of \citet{Wilms2000}, and we tie the hydrogen column density $N_\text{H}$ to be the same for all three spectra. The power law normalisation is left as a free parameter for each instrument to account for known discrepancies in absolute flux calibration. We tie the power law index between the \NuSTAR{} FPMA and FPMB, however allow it to be different for the EPIC-pn.

The top panel of Fig.~\ref{fig:X-calibration} shows the three spectra from the 16a epoch unfolded around the best-fitting absorbed power law model, and the bottom panel shows the data to (folded) model ratio. The EPIC-pn spectrum is slightly harder than the simultaneous \NuSTAR{} spectrum.  This is seen for all the observations, with the difference between the fitting power law indices listed in Table~\ref{tab:cross_calib}. We also see a discrepancy in the residuals at $\sim6$~keV corresponding to the \fekalpha{} line, with the line feature peaking at higher energies in EPIC-pn consistently across all of the epochs.
This discrepancy is most likely due to calibration issues with EPIC-pn timing mode. 
The EPIC-pn timing mode has previously been found to measure harder spectra than other observatories in addition to \NuSTAR{}. For example, an observation of GX 339-4 with a simultaneous \textit{RXTE} exposure \citep{Kolehmainen2014}, and an observation of the Rapid Burster in which \textit{Swift} XRT and \NuSTAR{} were in agreement, but both disagreed strongly with the EPIC-pn \citep{vandenEijnden2017}.

Here, we have used the most up to date (as of August 2022) EPIC-pn calibration files. Although they are an improvement on the old files, for example reducing the discrepancy in power law indices between the EPIC-pn and the \NuSTAR{} FPMs, problems do remain. We therefore
discard the EPIC-pn observations in this analysis.

\subsection{Timing Analysis}

\begin{figure}
    \centering
    \includegraphics{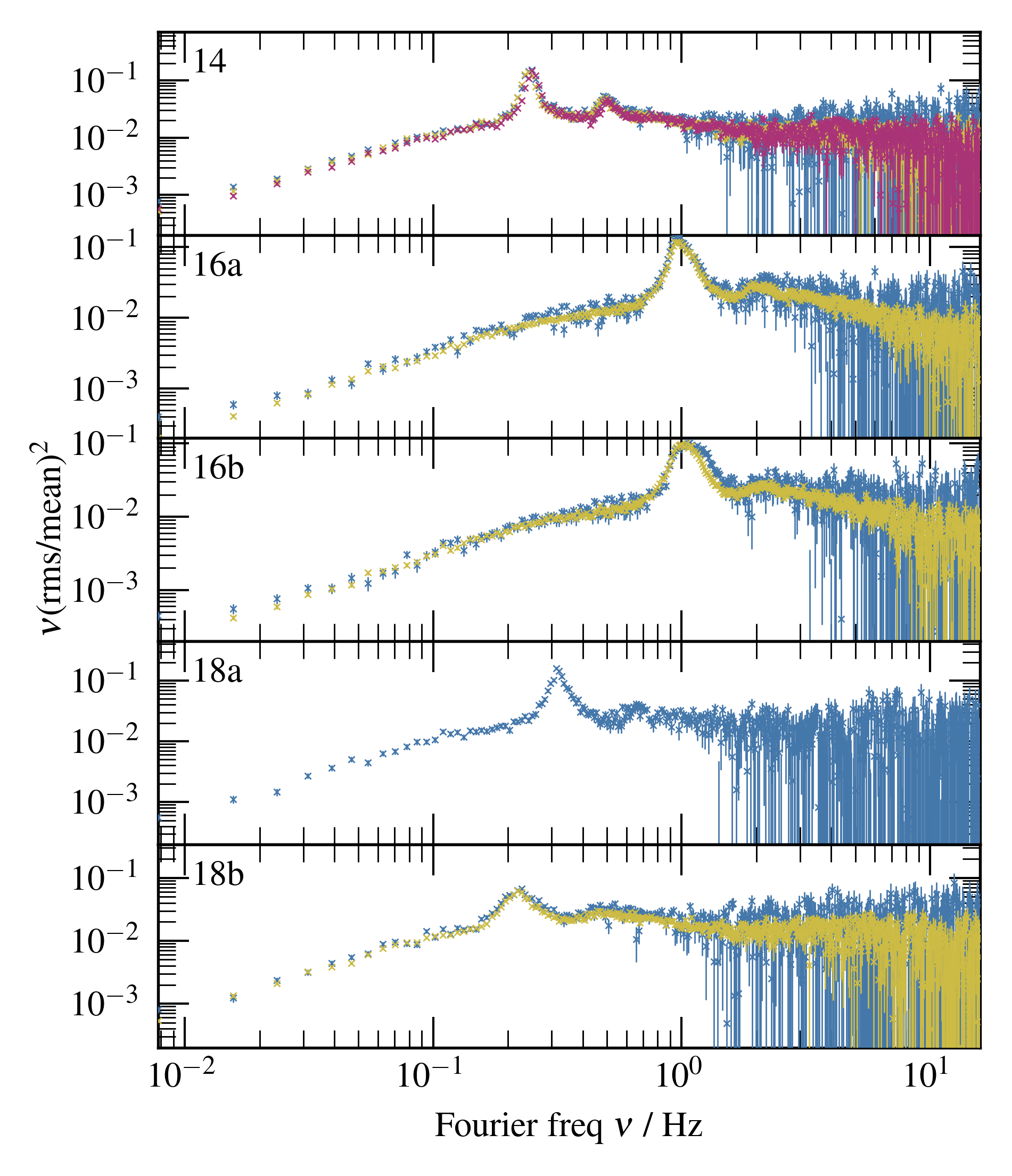}
    \caption{The power spectra of each epoch estimated for the \NuSTAR{} (blue) and \XMM{} (yellow and magenta) observations. In each observation, a strong type-C QPO is seen on top of the broad-band noise.  The QPOs span frequencies $0.2~\text{Hz}\lesssim\nu_\text{QPO}\lesssim1.2~\text{Hz}$, with a second harmonic at twice the fundamental.  }
    \label{fig:all_powerspectra}
\end{figure}

All the observations are from when H1743 was in the hard spectral state.  We briefly consider their timing properties by presenting the Poison noise-subtracted power-spectra of the 3--10~keV light curves of each observation in Fig.~\ref{fig:all_powerspectra}. For both the \NuSTAR{} and \XMM{} observations we used a time binning of $\delta t=1/32$~s, and ensemble averaged over $128$~s segments.  
For the \NuSTAR{} observations, the power-spectrum is estimated by the co-spectrum between the FMPA and FMPB lightcurves to correct deadtime as described in Section~2.4 of \citet{Nathan2022grs1915} using the technique of \citet{Bachetti2015}.
In each epoch, the estimate of the power spectrum from the \NuSTAR{} observation is shown in blue, with the \XMM{} observations shown in orange and green.  Slight differences in the shapes are likely caused by the observations not being completely simultaneous, or slightly different effective area curves of the instruments within the energy band.
Each epoch shows a strong type-C low-frequency QPO on top of the broad-band noise, with often a second harmonic showing.  The QPO frequency ranges from $\sim0.2$~Hz to $\sim1$~Hz.
The asymmetric shape of some of the QPOs (for example, 16b) also hints at frequency evolution of the QPO during the course of the observation.  Similarly, discrepancies between the QPOs seen with different telescopes (such as can be seen in epochs 14 and 16b) could be a result of this evolution as the joint observations are not strictly simultaneous, or possibly due to differences in the telescope responses even within the same energy band. These QPOs will be studied in greater detail in a forthcoming paper.

\section{Spectral Fits}
\label{sec:fits}
\subsection{The model}

\subsubsection{\textsc{rtdist}}
\label{sec:rtdist}
Here we use the version of \textsc{reltrans} which was presented in \citetalias{Ingram2022}, called \textsc{rtdist}.  
This model self-consistently calculates the ionisation parameter of the reflection spectrum, which allows the model to be sensitive to the ratio $M/D$.
In this section we summarise the key concepts of the time-averaged aspect of the model, which allows this sensitivity, but the interested reader should refer to \citetalias{Ingram2022} for the full details.
The direct continuum is represented by the model \textsc{NthComp} \citep{Zdziarski1996}, which considers Compton up-scattering of a thermal population of seed photons by a thermal population of electrons. The spectrum is a power law with photon index $\Gamma$ between low and high frequency breaks that are set respectively by the seed photon and electron temperature $kT_{\rm bb}$ and $kT_{\rm e}$. The restframe reflection spectrum is computed with \textsc{xillverDCp} \citep{Garcia2013,Garcia2016}, assuming an \textsc{NthComp} illuminating spectrum. The photon index and electron temperature are free parameters, whereas the seed photon temperature is set to $kT_{\rm bb}=0.05$ keV. \textsc{rtdist} accounts for general relativistic effects by tracing rays in the Kerr metric using the code \textsc{ynogk} \citep{Yang2013}. The corona is assumed to be a point-like source located on the BH spin axis a distance $h~\rg$ from the BH (i.e. the lamppost model). The angular emissivity of the corona is set by the parameters $b_1$, $b_2$ and the so-called boost parameter $\texttt{boost}$, defined such that increasing the boost parameter increases the amount of reflection with respect to the continuum. 
If all of these parameters are zero, the corona emits isotropically. For \texttt{boost}$>1$, emission is beamed towards the BH and for \texttt{boost}$<1$ is is beamed away from the BH. The parameters $b_1$ and $b_2$ influence the angular emissivity in a more subtle way (see \citetalias{Ingram2022} for details). The disc extends from an inner radius $r_{\rm in}$ to an outer radius $r_{\rm out}$ and has a constant scale height $h_d/r$. We freeze $r_{\rm out}$ to $1000~\rg$ and leave $r_{\rm in}$ as a free parameter. We freeze the BH spin parameter in our fits to $a_*=0.998$ to enable $r_{\rm in}$ to vary across the widest possible range without becoming smaller than the innermost stable circular orbit (ISCO) of the BH. The disc is assumed to be in the BH equatorial plane, and we view it from an inclination angle $i$.

The shape of the restframe reflection spectrum depends on the shape of the illuminating spectrum; the ionisation state of the disc; plus the electron density within the disc, $n_\text{e}$. 
Therefore, the reflection model \textsc{xillverDCp} takes as input parameters $\Gamma$, $kT_e$, the iron abundance relative to solar $A_{\rm Fe}$, $n_\text{e}$ and the ionisation parameter
\begin{equation}
    \xi = 4\pi\frac{F_\text{x}}{n_\text{e}},
    \label{eqn:xi}
\end{equation}
where $F_x$ is the illuminating X-ray flux. It has long been recognised that $\xi$ is a key parameter for determining ionisation balance in the disc \citep[e.g.][]{Ross2005}, but $n_\text{e}$ was limited to $n_e=10^{15}~{\rm cm}^{-3}$ in older reflection models. \cite{Garcia2016} showed that $n_\text{e}$ is also an important parameter in itself, and both $\xi$ and $n_\text{e}$ have since been simultaneously constrained with reflection spectroscopy for both AGN and XRBs \citep[see e.g.][]{Tomsick2018, Jiang2019a, Jiang2019b, Garcia2019, Liu2023}. As is theoretically expected \citep{Shakura1973}, $n_\text{e}$ is found to be larger for XRBs than for AGN. It is the simultaneous constraint of $\xi$ and $n_\text{e}$ that ultimately means that $F_x$ can be inferred. We can combine this flux with the knowledge of the emitting area from measuring $r_\text{in}/\rg{}$, to find the ratio of luminosity to BH mass squared $L_\text{R}/M^2$. 
The \emph{observed} reflected flux can then be compared to this inferred luminosity, thus making the observed reflection spectrum sensitive to the ratio $M/D$.

\textsc{rtdist} assumes a radial density profile\footnote{The model also provides the option to assume that $n_\text{e}$ is constant with radius, but here we use a radial density profile.} $n_e(r) \propto r^{3/2} \left(1-\sqrt{r_{\rm in}/r}\right)^{-2}$ \citep{Shakura1973} and takes as input parameters the normalisation of this relation, hereafter referred to as $n_\text{e}$, and the distance $D$. The ionisation parameter as a function of radius $\xi(r)$ is then calculated from input parameters including $D$, $M$ and $n_\text{e}$through \citep{Ingram2022}
\begin{equation}
    \xi(r) = \left(\frac{4\pi c^2 D}{G M}\right)^2 A \frac{\varepsilon(r)}{n_e(r)}\left(\frac{g_\text{sd}(r)}{g_\text{so}}\right)^{2-\Gamma}\int_\text{0.1 keV}^\text{1 MeV} f(E) \text{d}E\,,
    \label{eqn:ionisation_par}
\end{equation}
where $A$ is the model normalisation; $g_{sd}(r)$ and $g_{so}$ are the energy shifts for photons propagating from the source to radius $r$ on the disc, and from the source to the observer respectively; $f(E)$ is the illuminating flux spectrum; and $\varepsilon(r)$ is the emissivity profile.
The parameter $n_\text{e}$ is defined as $n_e(r)$ at the radius at which $\xi(r)$ reaches its peak value. We use a grid of 50 radial bins\footnote{The first 49 bins are logarithmically spaced between the disc inner radius $r_\text{in}$ and $300~\rg{}$, with the final bin encompassing everything up to our maximum disc outer radius $r_\text{out}=1000~\rg{}$.} for this calculation.

\subsubsection{Complete model}

The complete model we use is
$\textsc{const}\times\textsc{TBabs}\times\left(\textsc{diskbb}+\textsc{rtdist}+\textsc{xillverCp}\right)$.
In order, these components are used as follows:
\begin{itemize}
    \item $\textsc{const}$: We use a scalar value (which is constant with the spectral energy) to ensure an accurate absolute flux calibration.
    For \NuSTAR{}, this constant has been calibrated to within a $\sim 4\%$ systematic uncertainty via observations of the Crab nebula in `stray light' \citep{Madsen2017}, which is light that bypasses the mirror assembly but still impacts the detectors. As the data were reduced before the \NuSTAR{} calibration `20211020' was available, on the advice of the \NuSTAR{} SOC, we follow the analysis of \cite{Madsen2017} and use $\textsc{const}=0.86$ for the spectrum from \NuSTAR{}'s FPMA detector. We consider an uncertainty on this of $4\%$ when it comes to exploring parameter uncertainties. We leave this spectral constant as a free parameter for \NuSTAR{}'s FPMB and the \XMM{}'s RGS.
    \item \textsc{TBabs}: The effects of a line of sight absorption column, using the abundances of \citet{Wilms2000}.
    \item \textsc{diskbb}: This is used to model the intrinsic thermal disc spectrum.
    \item \textsc{rtdist}: The bulk of our model, described in Section~\ref{sec:rtdist}.
    \item \textsc{xillverCp}: A non-relativistic, distant reflector.%, with a spectrum unmodified by relativistic effects.
\end{itemize}

\subsection{Model fitting procedure}

Our strategy for model fitting considers \textit{individual fits}, in which we fit the model to each of our five epochs individually, and then a \textit{joint fit} when we fit the five epochs simultaneously, tying some key parameters to be the same for each epoch. For each of these, we first find a best fitting model using $\chi^2$ minimisation, before using Markov chain Monte Carlo (MCMC) sampling to explore parameter space. We use XSPEC version 12.12.0 throughout, for both $\chi^2$ minimisation and for running the MCMC. For all fits, we consider data from the \XMM{} RGS1 and RGS2 and the \NuSTAR{} FPMA and FPMB instruments.

\subsubsection{Individual fits}

In order to find an initial fit, we freeze $D=8.5$~kpc \citepalias{Steiner2012}. This is a sensible choice because the spectral model is only dependent on $M/D$, with a complete degeneracy between $M$ and $D$ (since we are not considering timing properties in this study). Therefore, we are effectively simplifying the model to have one model parameter 
$M_{8.5}=M\frac{8.5~\text{kpc}}{D}$
in place of the two parameters $M$ and $D$. We also freeze $\textsc{const}=0.86$ for the FPMA but leave it as a free parameter for FPMB, RGS1 and RGS2.
With the exception of $\textsc{const}$, the rest of the model parameters are kept the same for both the \NuSTAR{} and \XMM{} data. Epoch 14 includes two separate \textit{XMM-Newton} observations. We tie the parameters for these two observations to each other.

Our early analysis found that the model did not require the quadratic angular emissivity $b_1$, $b_2$ parameters, or the disc-scale height $h_d/r$ to be non-zero at anything other than very low significance, hence we freeze these to zero for our model fitting.  This corresponds to an emissivity profile modified from isotropic only by the `boost' parameter, and considering only a thin disc.  

We initialise the MCMC simulation from the best fitting model found via $\chi^2$ minimisation. 
We use an isotropic prior on the inclination, and a Jeffreys ($\propto1/p$) prior on the normalisations of the model components and $M_{8.5}$ in order to ensure that the parameter space is equally considered. 
At the end of the analysis we provide a distance to each step of the MCMC by sampling randomly from the marginalised MCMC distance posterior of \citetalias{Steiner2012}, and calculate the mass value associated with the value of $M_{8.5}$ for that step\footnote{The value of $8.5$~kpc comes from the mode of the distance distribution. Here we present parameters using the mean, so due to the skew of the \citetalias{Steiner2012} distance posterior this is closer to $8.1$~kpc.  This reporting does not effect our results as we use the entire \citetalias{Steiner2012} distance posterior, and will provide the full mass posterior for completeness.}.
Along with considering the distance, we also include the uncertainty on the FMPA calibration in a similar way, by assigning a new random value $\sim \mathcal{N}(\mu=0.86,\sigma=4\%)$ for the FMPA normalisation, and scaling the model normalisations to calculate a new mass value. As $M^2$ scales with the observed flux, this propagates a further $2\%$ uncertainty into $M$.
Our final posterior distribution on mass therefore includes uncertainty on the distance, as well as uncertainty on the absolute flux calibration of the instruments used.

\subsubsection{Joint fit}

We initialise our joint fit from the best fitting parameters of the individual fits. We then tie the mass $M_{8.5}$, inclination $i$, and iron abundance $A_\text{Fe}$ to be the same for all 5 epochs and find a new best fit by minimising $\chi^2$. All other previously free parameters remain free.
As with the individual fits, we then move to using MCMC.  Again, we allow the instrument normalisations to be free, with the same priors as before.

\subsection{Results of spectral fitting and MCMC}

\subsubsection{Individual fits}

\begin{figure*}
    \centering
    \includegraphics[width=\textwidth,trim={.5cm 1.5cm .5cm 1cm}]{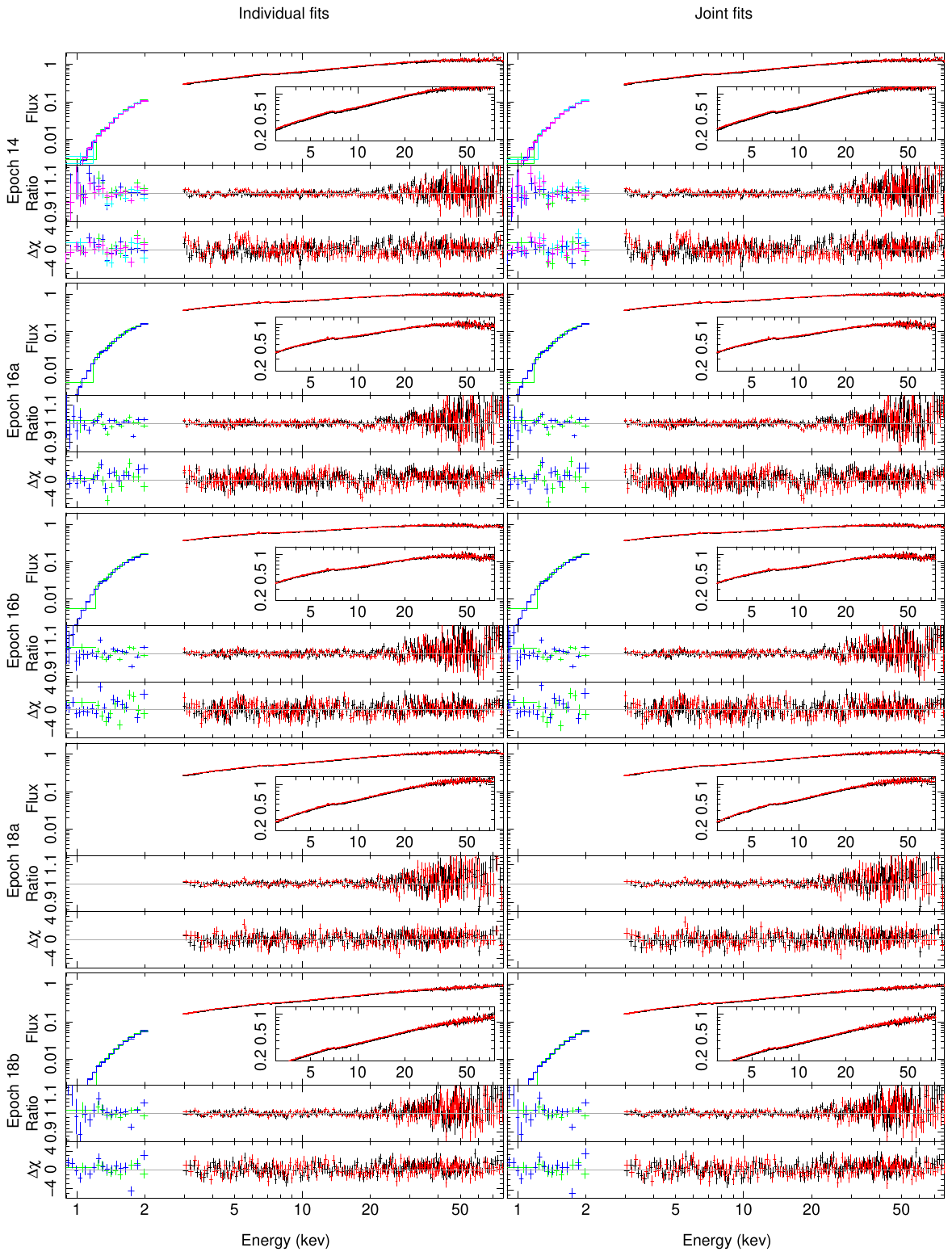}
    \caption{The observations and best fitting models unfolded around the telescope response matrices, plus the ratio of the data to the best fitting models and also the error-weighted residuals $\Delta\chi$.  The flux is given in units of $\text{keV}^2~\text{photons}/\text{s}/\text{cm}^2/\text{keV}$.
    On the left is the best fitting model to each epoch individually, while on the right is each epoch compared to the model found in the fit with the mass, inclination, and iron fraction tied.  For comparison, the axes all have the same scale.
    The \NuSTAR{} data are shown in black (FPMA) and red (FPMB), while \XMM{} data are shown in green, cyan (RGS1), and blue, magenta (RGS2). 
    The insets show a zoom-in of the $3-75$~keV region to show the \NuSTAR{} data more closely.}
    \label{fig:full_spectra_plot}
\end{figure*}

\begin{figure*}
    \centering
    \includegraphics{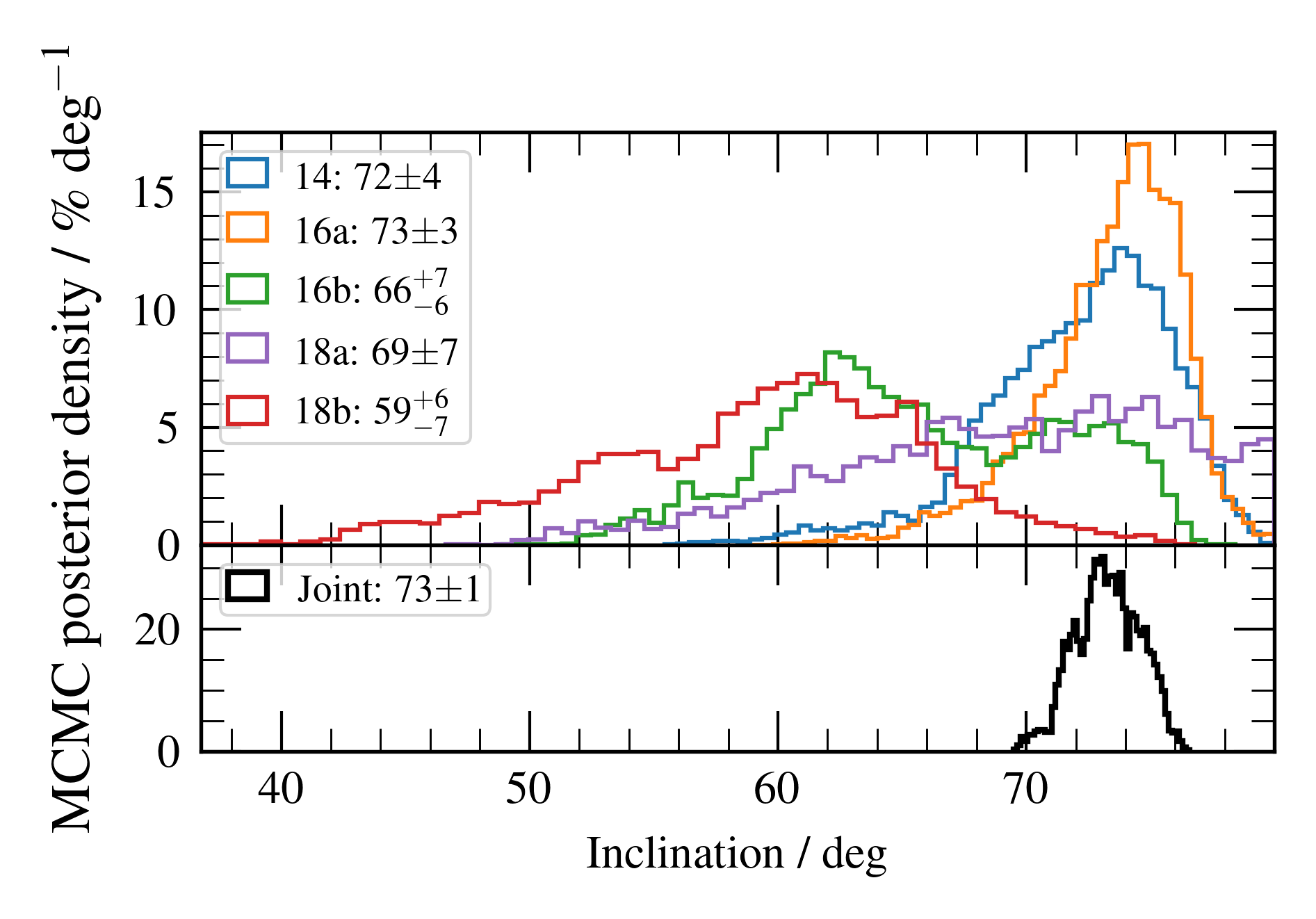}
    \includegraphics{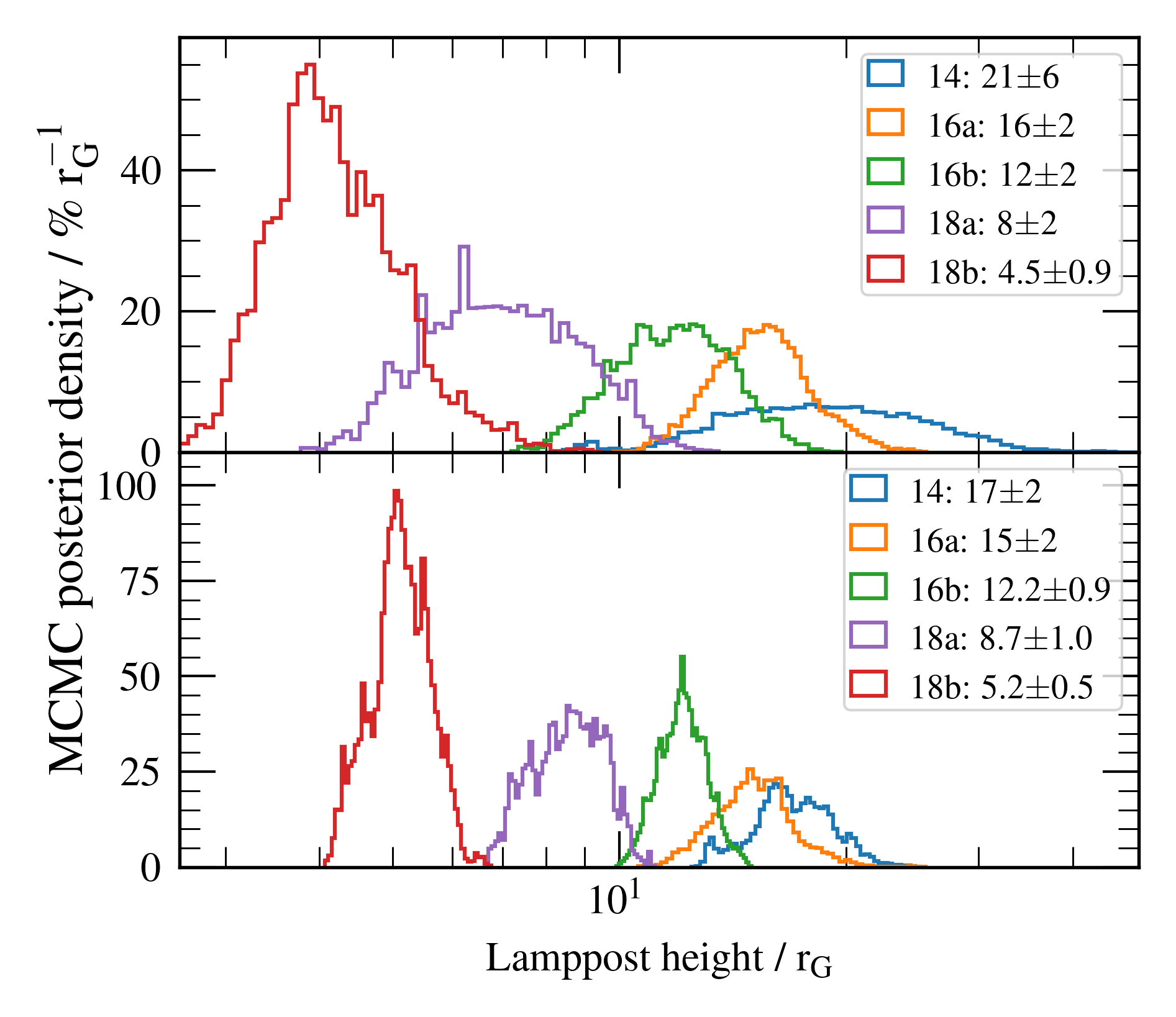}
    \includegraphics{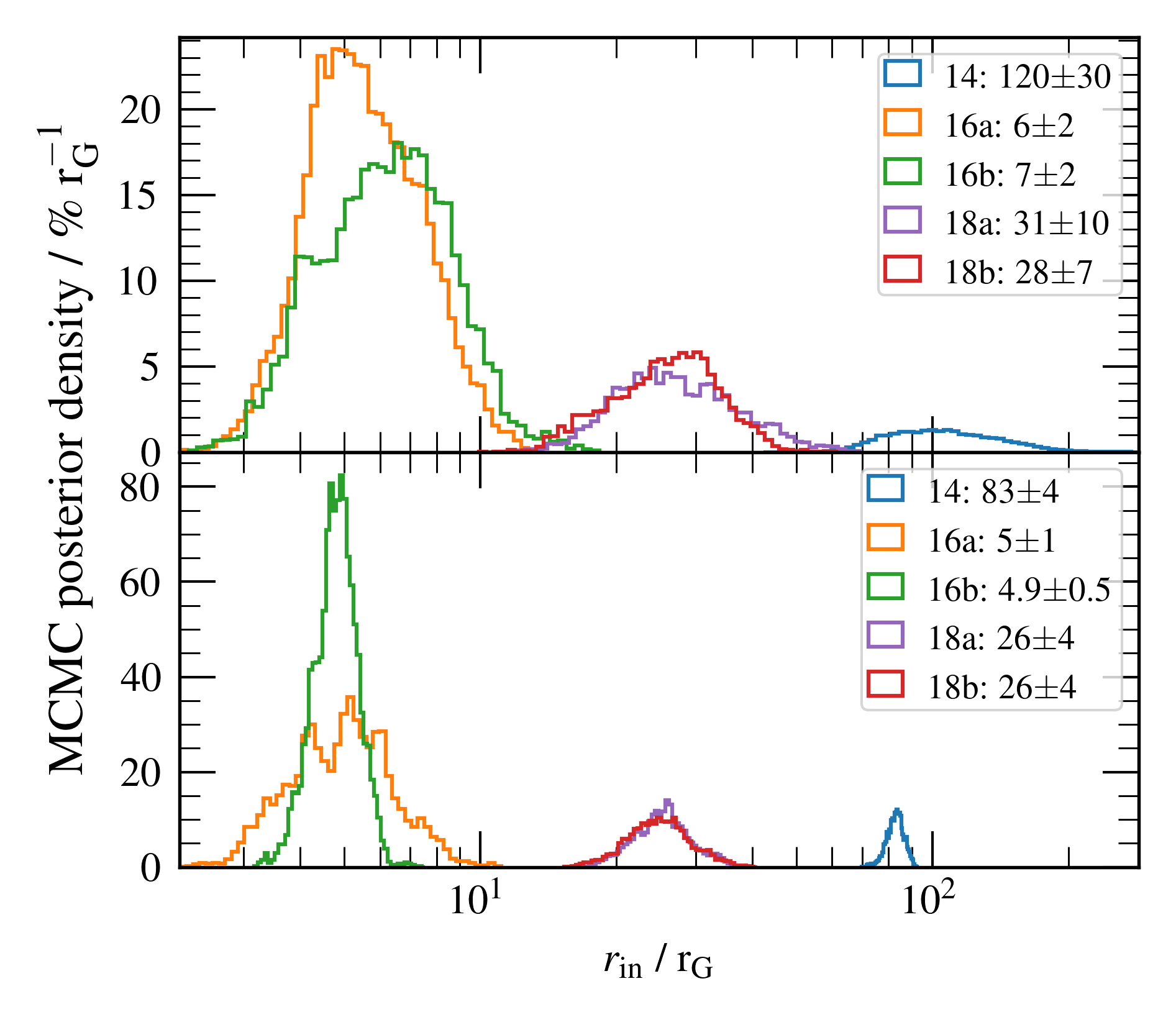}
    \includegraphics{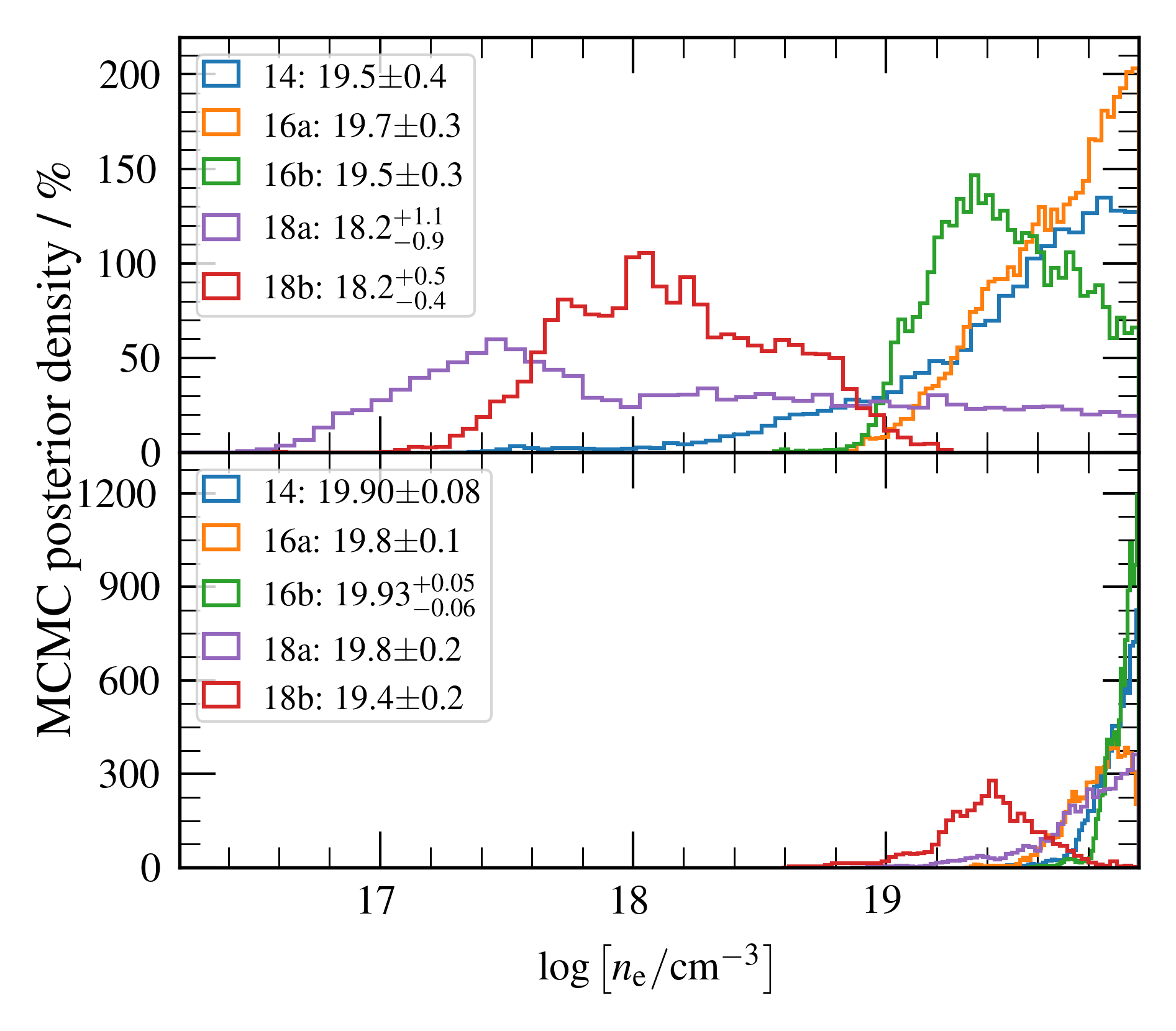}
    \caption{Histograms showing the MCMC posterior densities (in per cent) for the four parameters inclination, the height of the corona, disc inner radius, and disc electron density.  Each subfigure is split into an upper and lower panel.  The upper shows the results for the MCMCs run on the individual spectral fits, whereas the lower panel shows the result for the same parameters in the MCMC run on the joint spectral fit.  In each panel, the five epochs are colour-coded (14 in blue; 16a in orange; 16b in green; 18a in purple; and 18b in red).  As the inclination was tied between the epochs in the joint spectral fit, this is shown in black.  The legends list the posterior means along with the $1\sigma$ credible interval.  It should be noted that the model was capped with $\log N_\text{e}\leq 20$.}
    \label{fig:histograms}
\end{figure*}

\begin{figure}
    \centering
    %trim = {left bottom right top}
    \includegraphics[width=\columnwidth,trim={0 0.2cm 0 0},clip=true]{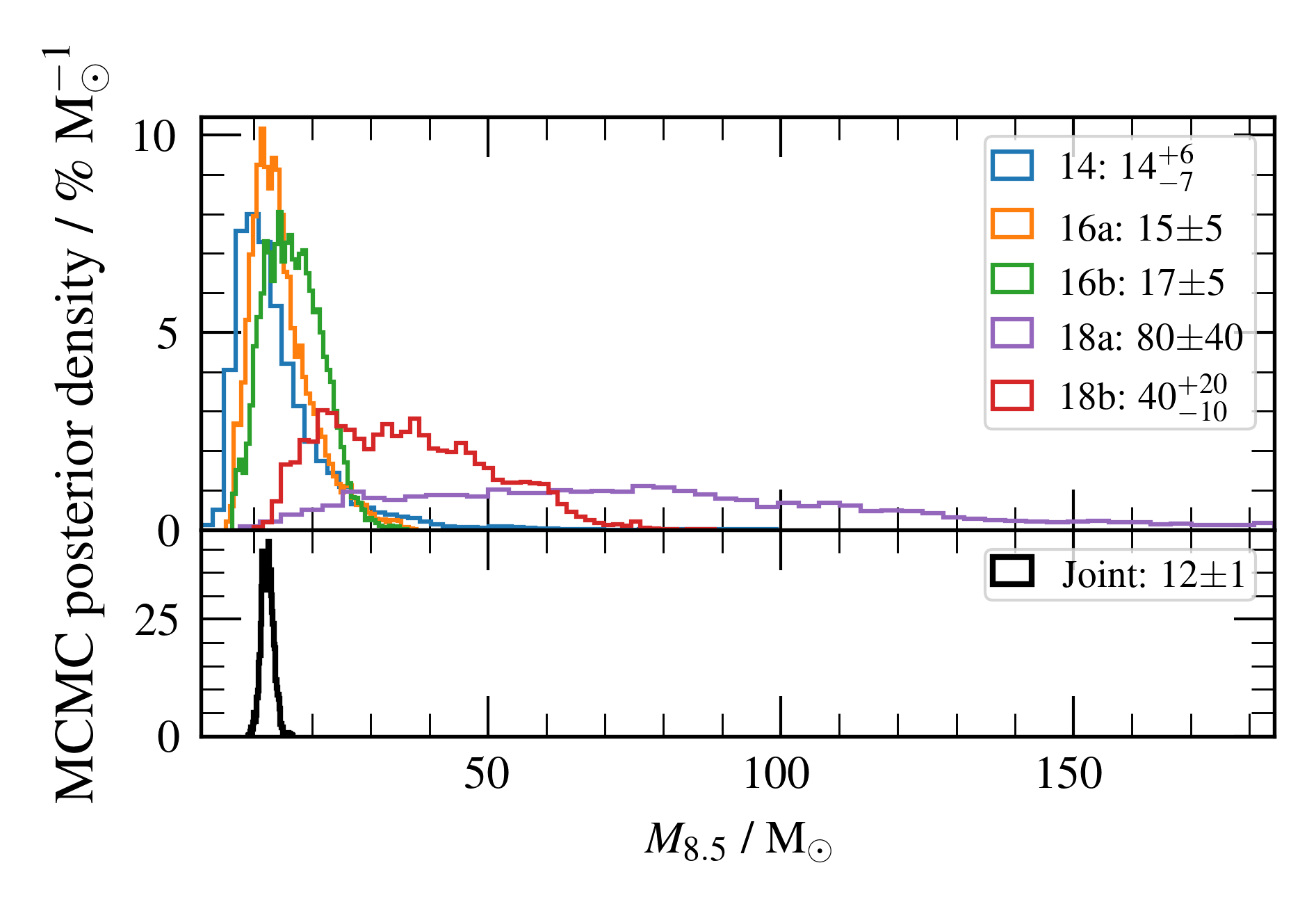}
    \includegraphics[width=\columnwidth,trim={0 0.2cm 0 0.2cm},clip=true]{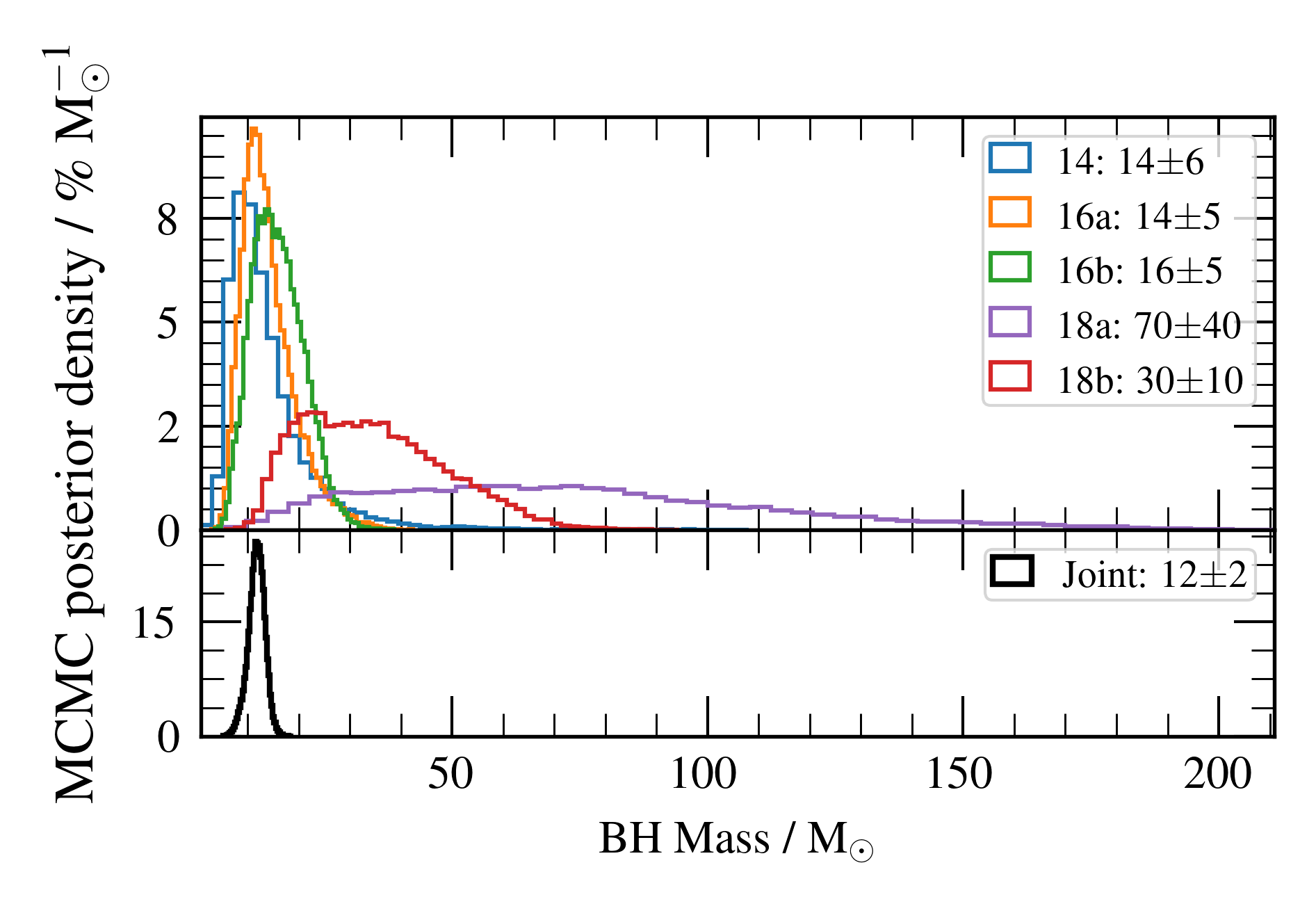} 
    \caption{The MCMC posterior density of the model parameter $M_{8.5}$, and then with this converted to the BH mass.  The histograms are stylistically the same as those in Fig.~\ref{fig:histograms}.  $M_{8.5}$ was the raw model parameter, from which the BH mass was calculated from $M_{8.5}$ by assigning each step within the MCMC a random distance and FPMA efficiency as described in the text.}
    \label{fig:mass_histograms}
\end{figure}

The spectra and the best fitting models of each epoch are shown on the left-hand side of Fig.~\ref{fig:full_spectra_plot}.
For each individual fit we ran a MCMC until convergence, and then ran $500$ walkers each for $7500$ steps.  We trimmed the MCMC by taking every $100^\text{th}$ step from each walker, to produce a final sample with $37500$ parameter combinations.  
From this, we show the posterior distributions of some key parameters, in the upper panels of the subplots in Figs.~\ref{fig:histograms} and \ref{fig:mass_histograms}, and the posterior mean and $1\sigma$ credible interval for each parameter is listed in Table~\ref{tab:individual_params}.

\begin{table*}
    \centering
    \begin{tabular}{lllllll}
    \toprule
    Component          & Variable                                           &    Epoch 14                              & Epoch 16a                               &   Epoch 16b                                        &  Epoch 18a                                     & Epoch 18b \\                                     
    \midrule
    \textsc{const}     & FMPB calibration                                   &    $0.8974\pm0.0009$                     & $0.8781\pm0.0008$                       &   $0.8785\pm0.0009$                                &  $0.8854\pm0.0013$                             & $0.8889^{+0.0014}_{-0.0013}$ \\                  
                       & RGS1 calibration                                   &    $0.671\pm0.009$                       & $0.75\pm0.01$                           &   $0.73\pm0.01$                                    &  \multicolumn{1}{c}{ --- }                     & $0.68\pm0.01$ \\                     
                       & RGS2 calibration                                   &    $0.646\pm0.009$                       & $0.73\pm0.01$                           &   $0.70\pm0.01$                                    &  \multicolumn{1}{c}{ --- }                     & $0.66\pm0.01$ \\                               
    \textsc{TBabs}     & $N_\text{H}$ / $10^{22}$ cm$^{-2}$                 &    $3.00\pm0.02$                         & $3.33\pm0.06$                           &   $3.10\pm0.07$                                    &  $3.16^{+0.14}_{-0.15}$                        & $3.16\pm0.04$ \\                                 
    \textsc{diskbb}    & $T_\text{in}$ / keV                                &    $0.83\pm0.14$                         & $0.201^{+0.007}_{-0.008}$               &   $0.224^{+0.015}_{-0.014}$                        &  $1.5\pm0.4$                                   & $0.091\pm0.007$ \\                               
                       & norm                                               &    $9^{+7}_{-6}$                         & $\left(8\pm3\right)\times10^{4}$        &   $\left(2.2^{+1.3}_{-1.2}\right)\times10^{4}$     &  $0.5^{+0.3}_{-0.4}$                           & $\left(2\pm2\right)\times10^{7}$\\               
    \textsc{rtdist}    & $M_{8.5}$ / $\Msun$                                &    $14^{+6}_{-7}$                        & $15\pm5$                                &   $17\pm5$                                         &  $80\pm40$                                     & $36\pm15$ \\                                     
                       & $M$ / $\Msun$                                      &    $14\pm6$                              & $14\pm5$                                &   $16\pm5$                                         &  $70\pm40$                                     & $35^{+15}_{-14}$ \\                              
                       & $i$ / degrees                                      &    $72\pm4$                              & $73\pm3$                                &   $66^{+7}_{-6}$                                   &  $69\pm7$                                      & $59^{+6}_{-7}$ \\                                
                       & $A_\text{Fe}$                                      &    $0.69\pm0.07$                         & $0.52\pm0.02$                           &   $0.54\pm0.03$                                    &  $0.68^{+0.15}_{-0.14}$                        & $0.72\pm0.07$ \\                                 
                       & $\Gamma$                                           &    $1.605\pm0.006$                       & $1.741\pm0.011$                         &   $1.729\pm0.012$                                  &  $1.642\pm0.014$                               & $1.582\pm0.011$ \\                               
                       & boost                                              &    $6\pm2$                               & $4\pm2$                                 &   $3\pm2$                                          &  $2.0^{+1.4}_{-1.3}$                           & $4\pm2$ \\                                       
                       & h / $\rg$                                          &    $21\pm6$                              & $16\pm2$                                &   $12\pm2$                                         &  $8\pm2$                                       & $4.5\pm0.9$ \\                                   
                       & $\text{k}T_\text{e}$ / keV                         &    $58\pm8$                              & $130\pm50$                              &   $290\pm150$                                      &  $39^{+7}_{-6}$                                & $330^{+120}_{-130}$ \\                           
                       & $\log_{10}[n_\text{e}~/~{\rm cm}^{-3}]$            &    $19.5\pm0.4$                          & $19.7\pm0.3$                            &   $19.5\pm0.3$                                     &  $18.2^{+1.1}_{-0.9}$                          & $18.2^{+0.5}_{-0.4}$ \\                          
                       & $r_\text{in}$ / $\rg$                              &    $120\pm30$                            & $6\pm2$                                 &   $7\pm2$                                          &  $30\pm10$                                     & $28\pm7$ \\                                      
                       & norm                                               &    $\left(1.5\pm0.4\right)\times10^{-2}$ & $\left(0.90\pm0.09\right)\times10^{-2}$ &   $\left(0.90^{+0.12}_{-0.11}\right)\times10^{-2}$ &  $\left(1.7\pm0.4\right)\times10^{-2}$         & $\left(4\pm2\right)\times10^{-2}$ \\             
    \textsc{xillverCP} & $\log_{10}[\xi~/~{\rm erg}~{\rm cm}~{\rm s}^{-1}]$ &    $3.75\pm0.05$                         & $0.6\pm0.4$                             &   $0.8^{+0.5}_{-0.6}$                              &  $1.2^{+0.8}_{-0.9}$                           & $1.1\pm0.8$ \\                                   
                       & norm                                               &    $\left(7\pm2\right)\times10^{-3}$     & $\left(1.8\pm0.5\right)\times10^{-3}$   &   $\left(0.9\pm0.4\right)\times10^{-3}$            &  $\left(0.9^{+0.5}_{-0.6}\right)\times10^{-3}$ & $\left(0.4^{+0.3}_{-0.4}\right)\times10^{-3}$ \\ 
    Fit statistic      & $\chi^2$ / DoF                                     &    3934 / 3612                           & 3562 / 3233                             &   3570 / 3212                                      &  2728 / 2722                                   & 3291 / 3151 \\
    \bottomrule
    \end{tabular}
    \caption{A summary of the parameters in the individual spectral fits, with the results provided by the MCMCs given as the posterior mean with the $1\sigma$ credible interval.  The calibration constants are given relative to the fixed FMPA calibration of $0.86$, and as described in the text the BH mass is calculated from $M_{8.5}$, with random sampling from a distance distribution and FPMA normalisation.}
    \label{tab:individual_params}
\end{table*}

\subsubsection{Joint fit}

The spectra of the best fitting joint model, with $M_{8.5}$, inclination and iron abundance tied across all five epochs, are shown
on the right side of Fig.~\ref{fig:full_spectra_plot}. 
For this joint fit we ran an MCMC until convergence, and then running $200$ walkers each for $52500$ steps.  We trimmed the MCMC by taking every $100^\text{th}$ step from each walker, to produce a final sample with $105000$ parameter combinations.  
The lower panels of the subplots in Figs.~\ref{fig:histograms} and \ref{fig:mass_histograms} show the posterior distributions of key parameters for the joint chain, including parameters which are tied between epochs, and those that are still free in each epoch.  Table~\ref{tab:joint_params} lists the posterior mean of each parameter, along with the the $1\sigma$ credible interval.

\begin{table*}
    \centering
    \begin{tabular}{lllllll}
    \toprule
    Component          & Variable                                             &	Epoch 14                                         &	Epoch 16a                             &	Epoch 16b                             &	Epoch 18a                                        &	Epoch 18b \\                                  
    \midrule
    \textsc{const}     & FMPB calibration                                     &	$0.8974^{+0.0007}_{-0.0008}$                     &	$0.878\pm0.001$                       &	$0.8786\pm0.0009$                     &	$0.8853^{+0.0013}_{-0.0012}$                     &	$0.8889^{+0.0011}_{-0.0012}$ \\               
                       & RGS1 calibration                                     &	$0.675\pm0.007$                                  &	$0.76\pm0.01$                         &	$0.740\pm0.007$                       &	\multicolumn{1}{c}{ --- }                        &	$0.677\pm0.014$ \\                            
                       & RGS2 calibration                                     &	$0.650\pm0.007$                                  &	$0.73\pm0.01$                         &	$0.709\pm0.006$                       &	\multicolumn{1}{c}{ --- }                        &	$0.652^{+0.013}_{-0.014}$ \\                  
    \textsc{TBabs}     & $N_\text{H}$ / $10^{22}$ cm$^{-2}$                   &	$2.99\pm0.02$                                    &	$3.35\pm0.06$                         &	$3.18\pm0.05$                         &	$3.43^{+0.15}_{-0.16}$                           &	$3.17\pm0.05$ \\                              
    \textsc{diskbb}    & $T_\text{in}$ / keV                                  &	$0.95\pm0.07$                                    &	$0.196\pm0.008$                       &	$0.207\pm0.007$                       &	$1.37^{+0.16}_{-0.15}$                           &	$0.0929^{+0.011}_{-0.008}$ \\                 
                       & norm                                                 &	$3.4^{+1.3}_{-1.1}$                              &	$\left(1.0\pm0.3\right)\times10^5$    &	$\left(4\pm1\right)\times10^4$        &	$1.1^{+0.4}_{-0.5}$                              &	$\left(1.8^{+1.6}_{-1.4}\right)\times10^7$ \\ 
    \textsc{rtdist}    & $M_{8.5}$ / $\Msun$                                  &	\multicolumn{5}{c}{ $12.4\pm1.0$ }               \\                                                                                                                                                                                 
                       & $M$ / $\Msun$                                        &	\multicolumn{5}{c}{ $12\pm2$ }                   \\                                                                                                                                                                                 
                       & $i$ / degrees                                        &	\multicolumn{5}{c}{	$73.3\pm1.4$ }               \\                                                                                                                                                                                 
                       & $A_\text{Fe}$                                        &	\multicolumn{5}{c}{	$0.54^{+0.02}_{-0.03}$ }     \\                                                                                                                                                                                 
                       & $\Gamma$                                             &	$1.610\pm0.004$                                  &	$1.752\pm0.007$                       &	$1.736\pm0.006$                       &	$1.631\pm0.006$                                  &	$1.605^{+0.007}_{-0.006}$ \\ 
                       & boost                                                &	$5.5^{+0.6}_{-0.5}$                              &	$1.1\pm0.4$                           &	$3.1^{+0.8}_{-0.7}$                   &	$0.6\pm0.2$                                      &	$1.9\pm0.6$ \\                                
                       & h / $\rg$                                            &	$17\pm2$                                         &	$15\pm2$                              &	$12.2\pm0.9$                          &	$9\pm1$                                          &	$5.2\pm0.5$ \\                                
                       & $\text{k}T_\text{e}$ / keV                           &	$59\pm7$                                         &	$80\pm20$                             &	$100\pm20$                            &	$34.4^{+1.4}_{-1.3}$                             &	$480\pm20$ \\                                 
                       & $\log_{10}[n_\text{e}~/~{\rm cm}^{-3}]$              &	$19.90\pm0.08$                                   &	$19.83\pm0.11$                        &	$19.93^{+0.05}_{-0.06}$               &	$19.8\pm0.2$                                     &	$19.4\pm0.2$ \\                               
                       & $r_\text{in}$ / $\rg$                                &	$83\pm4$                                         &	$5.4^{+1.3}_{-1.4}$                   &	$4.9\pm0.5$                           &	$26\pm4$                                         &	$26\pm4$ \\                                                   
                       & norm                                                 &	$\left(1.40^{+0.12}_{-0.13}\right)\times10^{-2}$ &	$\left(8.3\pm0.3\right)\times10^{-3}$ &	$\left(8.3\pm0.4\right)\times10^{-3}$ &	$\left(1.26^{+0.08}_{-0.07}\right)\times10^{-2}$ &	$\left(2.1\pm0.3\right)\times10^{-2}$ \\      
    \textsc{xillverCP} & $\log_{10}[\xi~/~{\rm erg}~{\rm cm}~{\rm s}^{-1}]$   &	$3.70^{+0.05}_{-0.04}$                           &	$0.17^{+0.15}_{-0.13}$                &	$0.5^{+0.5}_{-0.4}$                   &	$2.6^{+0.3}_{-0.4}$                              &	$1.8\pm0.3$ \\                                
                       & norm                                                 &	$\left(7.1\pm0.9\right)\times10^{-3}$            &	$\left(2.0\pm0.4\right)\times10^{-3}$ &	$\left(1.6\pm0.4\right)\times10^{-3}$ &	$\left(0.9\pm0.3\right)\times10^{-3}$            &	$\left(1.8\pm0.3\right)\times10^{-3}$   \\    
    Fit statistic      & $\chi^2$ / DoF                                       &	\multicolumn{5}{c}{	17096 / 15930 }     \\ 
    \bottomrule
    \end{tabular}
    
    \caption{A summary of the parameters in the join spectral fit, with the results provided by the MCMCs given as the posterior mean with the $1\sigma$ credible interval.  The calibration constants are given relative to the fixed FMPA calibration of $0.86$.  As described in the text the BH mass is calculated from $M_{8.5}$, with random sampling from a distance distribution and FPMA normalisation.  The parameters $M_{8.5}$ (and therefore BH mass), inclination, and $A_\text{Fe}$ are tied between the five epochs in this joint fit, so there is only one value for each of them.}
    \label{tab:joint_params}
\end{table*}

\subsubsection{Best fitting parameters}

Here we consider the overall viability of our best fitting parameters (as presented in Tables~\ref{tab:individual_params} and \ref{tab:joint_params}), before a detailed discussion in Section~\ref{sec:discussion}.   
We note that the RGS calibration constants are lower than the fixed FPMA constant of 0.86, as we saw for the EPIC-pn data in Table.~\ref{tab:cross_calib} -- although the values are not necessarily the same. As the absolute flux is well constrained by the \NuSTAR{} data, this calibration is not an issue, and the shape of the spectrum is able to constrain the absorption column.
We also notice that our disc density parameter is close to the model maximum of $\log_{10}[\xi~/~{\rm erg}~{\rm cm}~{\rm s}^{-1}]=20$, and we consider the implications of this in Section~\ref{sec:disc_discussion}. 
Further, as we will discuss in more detail in Section~\ref{sec:disc_discussion}, we note that the inner disc temperature $T_\text{in}$ covers a range of values (and the normalisation varies dependent on the temperature, although we do not tie this to physical units). While the temperature in epochs 14 and 18a seems higher than typically expected for hard state BH XRBs, further analysis demonstrates that while the inclusion of the component is statistically preferred, removing the \textsc{diskbb} does not change the key model results.
We note that our values for inner radius $r_\text{in}$ span a large range, however this is consistent with hard state X-ray binary behaviour.  Beyond these caveats, all of our parameters are reasonable and we therefore have confidence in our modelling.

\section{Revised Spin Measurement}
\label{sec:spin}

\begin{figure}
    \centering
    \includegraphics{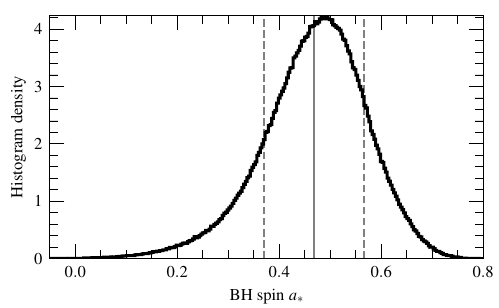}
    \caption{A histogram for the BH spin parameter $a_*$, applying the \citetalias{Steiner2012} approach but using our mass and inclination posteriors.  The vertical solid line is at the histogram mean, while the two vertical dashed lines are at the $\pm1\sigma$ quantiles.}
    \label{fig:JS_spin_hist}
\end{figure}

\citetalias{Steiner2012} used soft-state X-ray observations of H1743 (in which it can be assumed that $r_\text{in}=r_\text{ISCO}$) to estimate the BH spin $a_*$.  However, the method used requires a mass measurement.  As, at the time of publication no mass measurement was available, \citetalias{Steiner2012} used the \citet{Ozel2010} mass population distribution to estimate the BH spin parameter as $a_*=0.2\pm0.3$.  Our mass estimate of $12\pm2~\Msun{}$ is significantly higher than the $7.8~\Msun{}$ mean of the \citet{Ozel2010} distribution, and an examination of Fig.~3 of \citetalias{Steiner2012} suggests this should give a higher spin estimate.

We revisit the work of \citetalias{Steiner2012}, who fitted 170 \textit{Rossi X-ray Timing Explorer} \citep[\RXTE{};][]{Bradt1993} PCU2 observations 
with a Comptonised \textsc{KerrBB2} \citep{Li2005, McClintock2006} disc model (see Section~4 of \citetalias{Steiner2012}).
They created a logarithmically spaced grid over the BH $M$, $D$, and $\cos{i}$ parameters, and at each grid point fit all 170 observations with their model for two values of the viscosity parameter $\alpha=0.01\text{ or }0.1$ (allowing the Compton-up-scattered fraction of thermal photons $f_{SC}$, power law index $\Gamma$, spin parameter $a_*$, and mass accretion rate $\dot{M}$ to vary).  They applied a set of data selection criteria \citep[from ][]{Steiner2009} on the models at each vertex of their parameter grid to select sufficiently disc-dominated spectra and with reasonable luminosity.  For any grid-points reliant on poorer fit quality, the spin uncertainty was correspondingly increased.

We re-applied this method, updating the PCU2 calibration, keeping the 0.1\% systematic error, but using the \textsc{pcacorr} corrections \cite{Garcia2014pcacorr}. We employed an extended parameter grid, to which we match the values of $M$, $D$, and $\cos{i}$ from each step in our MCMC.  From the nearest grid vertex to each element in our MCMC chain, 5 spins are drawn from the corresponding distributions of $a_*$ for each of the two $\alpha$ settings.  These values are then combined into the histogram in Fig.~\ref{fig:JS_spin_hist}, which has a mean of $a_* = 0.47\pm0.10$ ($1\sigma$ uncertainty) . 

\section{Discussion}
\label{sec:discussion}

We have modelled the X-ray spectra of H1743 across 5 long-look observations with a reflection model that is sensitive to the ratio of BH mass to distance. By using the \citetalias{Steiner2012} distance posterior as a prior for our work, we obtain a mass measurement of $M=12\pm2~M_\odot$. This is amongst the first mass measurements of H1743, since extinction has prevented a direct dynamical mass measurement.
We then used this new mass estimate to revise a previous BH spin estimate obtained by modelling the thermal disc spectrum in archival soft state observations.
Here we discuss the implications of our results, along with caveats and suggested improvements to the model.

\subsection{Proof-of-principle}

This is the first study to measure BH mass from the reflection spectrum alone; although there have been previous attempts to measure mass from the X-ray spectrum. A crude estimate can be made by fitting the disc spectrum and assuming the inner radius to be at $6~r_g$ (e.g. \citealt{Zhang1997}, or for a review \citealt{McClintock2014}). \citet{Parker2016} improved on this concept by simultaneously fitting the disc and reflection components of GX~339-4, with the disc inner radius tied between the two, to yield estimates of the BH mass and the distance to the system. BH mass has also been estimated using the \textsc{TCAF} model \citep{Chakrabarti1995, Molla2017, Sreehari2019}, although there are typically several inconsistencies with these fits. For example, the \textsc{TCAF} model yields disc inner radii of hundreds of gravitational radii whereas the relativistically broadened iron line in the same observation favours disc inner radii commensurate with the ISCO.
\citet{Mastroserio2019} used an earlier version of \textsc{reltrans} to measure the mass of Cygnus~X-1 by simultaneously modelling the X-ray spectrum and timing properties. This X-ray reverberation mass measurement was slightly higher than the existing dynamical measurement \citep{Orosz2011}, but \cite{MillerJones2021} have since revised the optical measurement (based on a revised distance estimate) to agree remarkably well with \cite{Mastroserio2019}.

Here, we have used a newer version of \textsc{reltrans} \citepalias[\textsc{rtdist}; ][]{Ingram2022} to measure BH mass from the X-ray spectrum alone, without considering timing properties. The new model is sensitive to the ratio $M/D$ because the ionisation parameter used to calculate the restframe reflection spectrum is calculated self-consistently as a function of disc radius from the observed reflected flux, the disc density, the BH mass and the distance (see Eq.~\ref{eqn:ionisation_par}). We constrain the mass by inputting a previous measurement of the distance \citepalias{Steiner2012}); this study represents the first test of this method. It is encouraging that we obtain a reasonable mass measurement of $M=12\pm2~M_\odot$. 
We see our result as a proof-of-principle test of the validity of our assumptions, as the model is free to predict any BH mass.
Similar proof-of-principle tests were conducted by \citet{Shreeram2020}, \citet{Zdziarski2020}, and \citet{Connors2021} who modelled spectra of XRBs by fitting for the ionisation parameter, but then checked that their result was consistent with the ionisation parameter theoretically expected given the observed reflected flux, the measured disc density and the estimated BH mass and distance.

There are a number of simplifying assumptions in the model that could, in principle, have prevented it from returning a reasonable mass estimate. First, the \textsc{xillverDCp} model used to calculate the restframe reflection spectrum ignores the effect of intrinsic disc emission on the ionisation balance, instead assuming that the disc atmosphere is exclusively heated by the corona. This assumption more appropriate for lower disc temperatures, such as those associated with the hard state observations we analyse here. Second, the \textsc{xillverDCp} tables fix the seed photon temperature to $kT_{\rm bb} = 0.05$ keV, which is lower than the expected seed photon temperature for these observations, if they are provided by the disc as is commonly assumed. The broad-band shape of the illuminating spectrum influences the ionisation balance in the disc atmosphere, and thus the assumed value of $kT_{\rm bb}$ may influence the BH mass inferred from our method. Ideally, future versions of the model will have $kT_{\rm bb}$ as a free parameter, but it is encouraging that freezing it to a low value does not return a wildly unexpected result. 
Furthermore, the assumed \citet{Shakura1973} density profile could affect the the inferred total flux on the disc, and therefore our mass measurement in turn. However, the parameters $b_1$ and $b_2$ in the \textsc{rtdist} model have the ability to change the angular emissivity of the corona and therefore change the flux profile the disc. As we found that these parameters were consistent with zero, this is suggestive that the density profile assumption would not have a significant effect.
Finally, the radial disc emissivity profile in \textsc{reltrans} is calculated in the lamppost geometry, which is presumably a highly simplified representation of the true, extended corona (indeed X-ray recent polarimetric observations favour a radially extended corona in the case of Cygnus X-1; \citealt{Krawczynski2022}). Again, it is very encouraging that this assumption does not appear to prevent us from obtaining a reasonable result, especially given that we do not need to invoke the quadrature angular emissivity parameters $b_1$ and $b_2$.

In future work, we will conduct further, more stringent tests of the model by additionally considering X-ray timing properties. These tests will be particularly useful if we consider XRBs with existing mass and distance measurements, which can be compared with the values of these properties yielded by a full X-ray reverberation mapping analysis. Successful tests -- like this -- of the model bode well for further applications of this method, such as constraining the Hubble constant by measuring the distance to a sample of AGN \citepalias{Ingram2022}.

\subsection{Mass measurement}

We have estimated the mass of the BH in H1743 to be $12\pm2~\Msun{}$, which is consistent with other previous work \citep{Petri2008, Shaposhnikov2009, Ingram2014, Molla2017}, however unlike these previous estimates we have not made use of the QPOs.

We can consider this mass estimate in the context of \citet[][but also see \citealt{Farr2011}]{Ozel2010}, who used known BH mass measurements from 16 XRBs to fit an exponentially decaying mass function (which was motivated by theoretical expectations), and estimated the mass function
\begin{equation}
    P(M) = 
        \begin{cases}
            \frac{1}{1.57\Msun{}}\exp{\left[\frac{6.30\Msun{} - M}{1.57\Msun{}}\right]}, & M>6.3\Msun{}, \\
            0, & M\leq 6.3\Msun{}.
        \end{cases}
    \label{eqn:ozel_mass_dist}
\end{equation}
Our measured mass is towards the high end of this distribution. With the caveat that this work is primarily a proof-of-principle test of a new method, we do note that it is perhaps theoretically expected for a high-extinction source such as H1743 to have a larger BH mass than the low-extinction sources for which optical mass measurements have been possible \citep{Jonker2021}. This follows because XRB progenitor stars likely formed in the dense regions of the Galaxy, with some being ejected to their current high Galactic latitudes by a supernova kick \citep{Blaauw1961, Verbunt2017}. The prediction that the most massive BHs formed via direct collapse with no associated kick \citep{Fryer2001,Fryer2012} therefore implies that the XRBs that remained in dense regions after BH formation are systematically more massive than those that were ejected.  The method that we have presented here can be used to measure the masses of BHs in such systems.

\subsection{The new, higher spin measurement}

We have revised the \citetalias{Steiner2012} BH spin parameter estimate using our results, from $a_*=0.2\pm0.3$ to $a_*=0.47\pm0.10$.  There have been a few attempts to constrain the BH spin using QPOs before.
Our measurement is consistent with the measurement $0.2\leq a_*\leq0.6$ by \citet{Tursunov2018}, likewise with the constraint $a_* \geq 0.21$, although we disagree with $a_* > 0.68$ from \citet{Mondal2010}.
Rather than using methods involving QPO frequencies, \citet{Draghis2023} used reflection modelling of \NuSTAR{} observations (including the data used in this work) with variants of the \textsc{relxill} models \citep{Garcia2014}, and found a spin parameter $a_*=0.98^{+0.01}_{-0.02}$.  This is not only inconsistent with our value of the spin parameter, but many of our fit parameters disagree, including the source inclination -- our $73.3\pm1.4^\circ$ inclination  is consistent with the $75\pm3^\circ$ radio jet inclination of \citetalias{Steiner2012}, whereas \citet{Draghis2023} found a lower $56\pm4^\circ$ inclination.  However, while the work in this paper does involve reflection modelling, the value of the spin parameter we provide uses the continuum-fitting method and data employed in \citetalias{Steiner2012}, whereas \citet{Draghis2023} uses a pure reflection-modelling technique.  
We also note that our value for the spin corresponds to an ISCO located at $4.3\rg$; the values of our reflection modelling finds for the inner disk radii $r_\text{in}$ are consistent with this measurement, as can be seen in Fig.~\ref{fig:rin_QPO_freq_plot} which shows our $r_\text{in}$ measurements and the values which would be excluded the ISCO based on our spin measurement.

\subsection{Inner radius, inclination, and QPOs}

\begin{figure}
    \centering
    \includegraphics[width=\columnwidth,trim={0.2cm 0.2cm 0 0.2cm},clip=true]{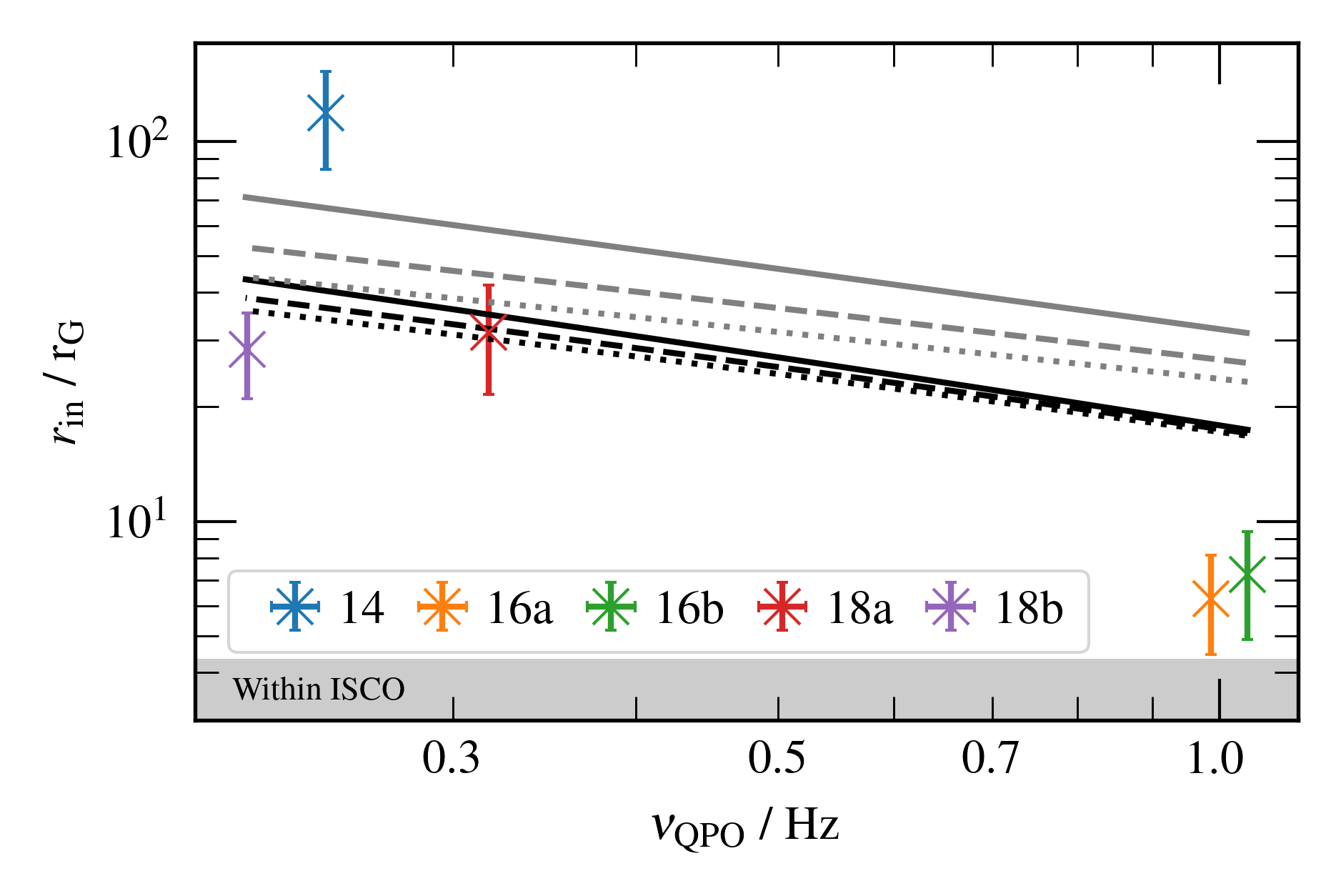}
    \caption{The disc inner radius (mean and $1\sigma$ credible interval) of each epoch from the MCMC of the joint fit, as compared to the QPO frequency measured in the $3-10$~keV band (see Fig.~\ref{fig:all_powerspectra}).  The lines are examples of the QPO frequency vs disc inner radius from Eq.~2 of \citet{Ingram2009}, using our best fit mass of $12~\Msun$ and our spin $a_*=0.47$.  The black lines have the precessing flow's inner radius set by the bending radius  with $h/r=0.15$, whereas the grey lines show the flow down to the ISCO.  Three different surface density power laws were used: $\zeta=0.5$ (solid line); $\zeta=0.0$ (dashed line); and $\zeta=-0.45$ (dotted line). The shaded region of the plot indicates inner disc radii which would lie within the ISCO.}
    \label{fig:rin_QPO_freq_plot}
\end{figure}

In Fig.~\ref{fig:all_powerspectra} we demonstrate the type-C QPO seen in each of the observational epochs, spanning the frequency range $0.2~\text{Hz} \lesssim \nu_\text{QPO} \lesssim 1.1~\text{Hz}$.  In Fig.~\ref{fig:rin_QPO_freq_plot} we compare the QPO frequency in each epoch with our measured disc inner radius, $r_\text{in}$ (from the joint fit). We see that %, although the error bars are large, 
the inner radius is consistent with reducing as the QPO frequency increases, as is predicted for many of QPO models \citep{Ingram2019review}. We additionally plot this predicted relationship for the precessing inner flow model of \cite{Ingram2009}. In this model, the entire corona precesses due to the frame dragging effect. The precession frequency depends on the BH mass and spin, the inner radius of the corona, the outer radius of the corona (assumed to be the inner radius of the disc) and the power law index $\zeta$ characterising the surface density profile within the corona, $\Sigma(r) \propto r^{-\zeta}$. 
We calculate the precession frequency for a range of disc inner radii by fixing $M$ to our best fitting value, the spin to $a_*=0.47$ as implied by our new mass estimate, and each line represents a different assumption about the corona inner radius and $\zeta$.  We consider three values for $\zeta$: $\zeta=0.5$ (solid lines) from classical advection-dominated accretion flows; $\zeta=0.0$ (dashed lines) as approximately seen in numerical simulations; and $\zeta=-0.45$ (dotted lines) close to a thin disc.  The corona inner radius was either taken to be the ISCO of the BH (black lines) or the bending wave radius for a disc scale height $h_d/r=0.15$ (grey lines; \citealt{Lubow2002,Ingram2009}).

We see that the models all reproduce the reduction in disc inner radius with QPO frequency, as expected, but our measured $r_{\rm in}$ values for the higher QPO frequencies we sample ($\sim1$~Hz) are smaller than the model predictions -- as is common for reflection modelling of bright hard state and intermediate state observations \citep[e.g.][]{Garcia2015,Fuerst2016}. This could of course be due to the precession model being incorrect or incomplete, or could instead be due to systematic errors associated with reflection modelling. For example, here we represent the corona as a single temperature region, whereas an extended, stratified corona would be better represented by a sum of Comptonised spectra each with different power law indices and high energy cut offs. The resulting curvature of the continuum spectrum typically leaves a weaker red wing in the iron line residuals, thus indicating a larger disc inner radius \citep[e.g.][]{Basak2017,Zdziarski2022}. 
These systematic effects in the reflection modelling could be represented by the scatter between the $r_\text{in}$ measurements between observations, as seen in Fig.~\ref{fig:rin_QPO_freq_plot}.
Alternatively the discrepancy could be because the formula we have used for the precession frequency ignores the torque from the outer disc. Since this torque acts to slow down precession, accounting for it would decrease the disc inner radius inferred from the precession model. This effect has recently been seen in the first numerical simulations of a truncated disc / hot inner flow geometry misaligned with the BH spin axis \citep{Bollimpalli2023}.

The precessing inner flow model of \cite{Ingram2009} for the QPOs requires a moderate misalignment ($\sim 15^\circ$) between the inner flow of the disc and the BH spin axis -- which we assume to be aligned with the jet.
From the joint fit, we find a disc inclination of $[73.3\pm1.4]^\circ$, which is consistent with the ballistic jet inclination of $[75\pm 3]^\circ$ found by \citetalias{Steiner2012}, but does not preclude a $\sim 15^\circ$ misalignment (given the unknown projection of the binary orbital axis on the sky; \citealt{Ingram2017h1743}) required for the QPOs within this model. 

\subsection{Density and the disc}
\label{sec:disc_discussion}
We find that a high disc density is required in our fits, particularly in the joint fit (Fig.~\ref{fig:histograms}). This is theoretically expected for an XRB ($n_e \propto 1/M$ for a radiation pressure dominated disc; \citealt{Shakura1973}). However, we are unable to sample the high density end of the $n_\text{e}$ posterior due to the \textsc{xillverDCp} grid having a maximum density of $n_\text{e}=10^{20}~\text{cm}^{-3}$. This is due to numerical issues associated with larger densities. Within the \textsc{rtdist} model, there is a correlation between the parameters $M \sim 1/\sqrt{n_e}$ and so our inability to sample the highest densities means that we very likely under predict the low mass region of the mass posterior.  
Further, as our density profile corresponds to the lowest density of the \citet{Shakura1973} radial profile we use, this has the effect of flattening the profile to be close to being constant.
This model systematic could potentially be important, and our result has to be considered in the context of the particular reflection models we have used, including all the relevant assumptions these models include. We cannot investigate higher density until the \textsc{xillver} reflection tables are extended to these densities, however the statistical uncertainty of our result demonstrates how this method could be used when using future versions of reflection models.
\citet{Liu2023} studied a number of observations of different BH XRBs, including the same \NuSTAR{} observations we use here.  
They considered a version of the \textsc{reflionx} refection model which has a maximum electron density higher than that of our model. Of the observations that both we and they have considered, the highest electron density they report is $\log_{10}[n_\text{e}~/~{\rm cm}^{-3}] = 20.9^{+0.2}_{-0.7}$; this observation is ObsID 90401335002 -- our epoch 18a -- which we found to have a poorly constrained density when we modelled the spectrum individually.  The other values they report are consistent with our disc densities.

We note that we see a range of inner disc temperatures $T_\text{in}$, particularly for epochs 14 and 18a which our modelling finds temperatures higher than typically expected for hard state BH XRBs.  Within the joint fit, removing the disc component for these two epochs does not significantly change the key model results, albeit causing an unacceptable change in the fit statistic.  Removing the disc component from the model entirely caused the disc density to peg against the hard limit $\log_{10}[n_\text{e}~/~{\rm cm}^{-3}]=20$, as reflection models with a higher disc density contain more flux at lower energies.  In the \textsc{rtdist} model, a higher disc density correlates with a lower mass, thus it is of no surprise that this higher density reduces our mass estimate to $M_{8.5}=9.4^{+1.5}_{-0.8}$.  This lower mass estimate is, however, still strongly statistically unpreferred -- and we note that this change in the model assumptions only causes a small change in our result.

\section{Conclusions}
\label{sec:conclusions}

Here we have presented a mass-measurement of the BH in the XRB H1743-322, among the first BH mass measurements to use the X-ray spectrum alone.  By using the X-ray reflection spectrum shape to infer the X-ray flux incident onto (and therefore reflected from) the disc, and comparing that to the flux we observe, we were able to constrain the BH mass-distance ratio $M/D$.  Past measurements of the source distance using the radio jets by \citetalias{Steiner2012} gave a distance of $D=8.5 \pm 0.8$~kpc.  We used our model to measure $M_{8.5}$, the BH mass if the source was 8.5~kpc from us, before folding in the distance measurement and uncertainty to find $M=12\pm2~\Msun$. This provides an excellent proof of principle of this technique; although we note that maximum limit on the disc electron density within the current \textsc{xillver} models could bias our result against lower BH masses.

Using MCMC analysis we found a posterior distribution for the BH mass, and we were able to use the same methodology as \citetalias{Steiner2012} to calculate the BH spin based upon our measurement.  As our mass posterior is higher than the mean of the Galactic XRB mass distribution, which had previously been used, our X-ray continuum-fitting spin measurement of $a_*=0.47\pm0.10$ is higher than that previously found by \citetalias{Steiner2012}.  

\section*{Acknowledgements}
The authors thank the referee for their helpful comments. 
The authors also thank Paul Tol for his publicly available, colour-blind accessible colour schemes which are used in Fig.~\ref{fig:all_powerspectra}. 
E. N. and A. I. acknowledge support from the Royal Society.
E. N. acknowledges support from NASA theory grant 80NSSC20K0540.
This research has made use of data and/or software provided by the High Energy Astrophysics Science Archive Research Center (HEASARC), which is a service of the Astrophysics Science Division at NASA/GSFC.

%%%%%%%%%%%%%%%%%%%%%%%%%%%%%%%%%%%%%%%%%%%%%%%%%%
\section*{Data Availability}

%The inclusion of a Data Availability Statement is a requirement for articles published in MNRAS. Data Availability Statements provide a standardised format for readers to understand the availability of data underlying the research results described in the article. The statement may refer to original data generated in the course of the study or to third-party data analysed in the article. The statement should describe and provide means of access, where possible, by linking to the data or providing the required accession numbers for the relevant databases or DOIs.
The observational data used in this research are public and available for download from the HEASARC. The \textsc{reltrans} model package is publicly available, and the public version will soon be updated to include the \textsc{rtdist} model used in this paper.

%%%%%%%%%%%%%%%%%%%% REFERENCES %%%%%%%%%%%%%%%%%%

% The best way to enter references is to use BibTeX:

\bibliographystyle{mnras}
\bibliography{bib}

\begin{thebibliography}{}
\makeatletter
\relax
\def\mn@urlcharsother{\let\do\@makeother \do\$\do\&\do\#\do\^\do\_\do\%\do\~}
\def\mn@doi{\begingroup\mn@urlcharsother \@ifnextchar [ {\mn@doi@}
  {\mn@doi@[]}}
\def\mn@doi@[#1]#2{\def\@tempa{#1}\ifx\@tempa\@empty \href
  {http://dx.doi.org/#2} {doi:#2}\else \href {http://dx.doi.org/#2} {#1}\fi
  \endgroup}
\def\mn@eprint#1#2{\mn@eprint@#1:#2::\@nil}
\def\mn@eprint@arXiv#1{\href {http://arxiv.org/abs/#1} {{\tt arXiv:#1}}}
\def\mn@eprint@dblp#1{\href {http://dblp.uni-trier.de/rec/bibtex/#1.xml}
  {dblp:#1}}
\def\mn@eprint@#1:#2:#3:#4\@nil{\def\@tempa {#1}\def\@tempb {#2}\def\@tempc
  {#3}\ifx \@tempc \@empty \let \@tempc \@tempb \let \@tempb \@tempa \fi \ifx
  \@tempb \@empty \def\@tempb {arXiv}\fi \@ifundefined
  {mn@eprint@\@tempb}{\@tempb:\@tempc}{\expandafter \expandafter \csname
  mn@eprint@\@tempb\endcsname \expandafter{\@tempc}}}

\bibitem[\protect\citeauthoryear{{Abbott} et~al.,}{{Abbott}
  et~al.}{2021}]{Abbott2021}
{Abbott} R.,  et~al., 2021, \mn@doi [\apjl] {10.3847/2041-8213/abe949}, \href
  {https://ui.adsabs.harvard.edu/abs/2021ApJ...913L...7A} {913, L7}

\bibitem[\protect\citeauthoryear{{Bachetti} et~al.,}{{Bachetti}
  et~al.}{2015}]{Bachetti2015}
{Bachetti} M.,  et~al., 2015, \mn@doi [\apj] {10.1088/0004-637X/800/2/109},
  \href {https://ui.adsabs.harvard.edu/abs/2015ApJ...800..109B} {800, 109}

\bibitem[\protect\citeauthoryear{{Bambi} et~al.,}{{Bambi}
  et~al.}{2021}]{Bambi2021}
{Bambi} C.,  et~al., 2021, \mn@doi [\ssr] {10.1007/s11214-021-00841-8}, \href
  {https://ui.adsabs.harvard.edu/abs/2021SSRv..217...65B} {217, 65}

\bibitem[\protect\citeauthoryear{{Basak}, {Zdziarski}, {Parker}  \&
  {Islam}}{{Basak} et~al.}{2017}]{Basak2017}
{Basak} R.,  {Zdziarski} A.~A.,  {Parker} M.,   {Islam} N.,  2017, \mn@doi
  [\mnras] {10.1093/mnras/stx2283}, \href
  {https://ui.adsabs.harvard.edu/abs/2017MNRAS.472.4220B} {472, 4220}

\bibitem[\protect\citeauthoryear{{Belloni}}{{Belloni}}{2010}]{Belloni2010}
{Belloni} T.~M.,  2010, in {Belloni} T.,  ed., , Vol.~794, Lecture Notes in
  Physics, Berlin Springer Verlag.
Springer Berlin Heidelberg, pp 53--84, \mn@doi{10.1007/978-3-540-76937-8\_3}

\bibitem[\protect\citeauthoryear{{Blaauw}}{{Blaauw}}{1961}]{Blaauw1961}
{Blaauw} A.,  1961, \bain, \href
  {https://ui.adsabs.harvard.edu/abs/1961BAN....15..265B} {15, 265}

\bibitem[\protect\citeauthoryear{{Bollimpalli}, {Fragile}  \&
  {Klu{\'z}niak}}{{Bollimpalli} et~al.}{2023}]{Bollimpalli2023}
{Bollimpalli} D.~A.,  {Fragile} P.~C.,   {Klu{\'z}niak} W.,  2023, \mn@doi
  [\mnras] {10.1093/mnrasl/slac155}, \href
  {https://ui.adsabs.harvard.edu/abs/2023MNRAS.520L..79B} {520, L79}

\bibitem[\protect\citeauthoryear{{Bradt}, {Rothschild}  \& {Swank}}{{Bradt}
  et~al.}{1993}]{Bradt1993}
{Bradt} H.~V.,  {Rothschild} R.~E.,   {Swank} J.~H.,  1993, \aaps, \href
  {https://ui.adsabs.harvard.edu/abs/1993A&AS...97..355B} {97, 355}

\bibitem[\protect\citeauthoryear{{Casares} \& {Jonker}}{{Casares} \&
  {Jonker}}{2014}]{Casares2014}
{Casares} J.,  {Jonker} P.~G.,  2014, \mn@doi [\ssr]
  {10.1007/s11214-013-0030-6}, \href
  {https://ui.adsabs.harvard.edu/abs/2014SSRv..183..223C} {183, 223}

\bibitem[\protect\citeauthoryear{{Chakrabarti} \& {Titarchuk}}{{Chakrabarti} \&
  {Titarchuk}}{1995}]{Chakrabarti1995}
{Chakrabarti} S.,  {Titarchuk} L.~G.,  1995, \mn@doi [\apj] {10.1086/176610},
  \href {https://ui.adsabs.harvard.edu/abs/1995ApJ...455..623C} {455, 623}

\bibitem[\protect\citeauthoryear{{Connors} et~al.,}{{Connors}
  et~al.}{2021}]{Connors2021}
{Connors} R. M.~T.,  et~al., 2021, \mn@doi [\apj] {10.3847/1538-4357/abdd2c},
  \href {https://ui.adsabs.harvard.edu/abs/2021ApJ...909..146C} {909, 146}

\bibitem[\protect\citeauthoryear{{Corral-Santana}, {Casares},
  {Mu{\~n}oz-Darias}, {Bauer}, {Mart{\'\i}nez-Pais}  \&
  {Russell}}{{Corral-Santana} et~al.}{2016}]{Corral-Santana2016}
{Corral-Santana} J.~M.,  {Casares} J.,  {Mu{\~n}oz-Darias} T.,  {Bauer} F.~E.,
  {Mart{\'\i}nez-Pais} I.~G.,   {Russell} D.~M.,  2016, \mn@doi [\aap]
  {10.1051/0004-6361/201527130}, \href
  {https://ui.adsabs.harvard.edu/abs/2016A&A...587A..61C} {587, A61}

\bibitem[\protect\citeauthoryear{{Dauser}, {Garcia}, {Wilms}, {B{\"o}ck},
  {Brenneman}, {Falanga}, {Fukumura}  \& {Reynolds}}{{Dauser}
  et~al.}{2013}]{Dauser2013}
{Dauser} T.,  {Garcia} J.,  {Wilms} J.,  {B{\"o}ck} M.,  {Brenneman} L.~W.,
  {Falanga} M.,  {Fukumura} K.,   {Reynolds} C.~S.,  2013, \mn@doi [\mnras]
  {10.1093/mnras/sts710}, \href
  {https://ui.adsabs.harvard.edu/abs/2013MNRAS.430.1694D} {430, 1694}

\bibitem[\protect\citeauthoryear{{Done}, {Gierli{\'n}ski}  \& {Kubota}}{{Done}
  et~al.}{2007}]{Done2007}
{Done} C.,  {Gierli{\'n}ski} M.,   {Kubota} A.,  2007, \mn@doi [\aapr]
  {10.1007/s00159-007-0006-1}, \href
  {https://ui.adsabs.harvard.edu/abs/2007A&ARv..15....1D} {15, 1}

\bibitem[\protect\citeauthoryear{{Draghis}, {Miller}, {Zoghbi}, {Reynolds},
  {Costantini}, {Gallo}  \& {Tomsick}}{{Draghis} et~al.}{2023}]{Draghis2023}
{Draghis} P.~A.,  {Miller} J.~M.,  {Zoghbi} A.,  {Reynolds} M.,  {Costantini}
  E.,  {Gallo} L.~C.,   {Tomsick} J.~A.,  2023, \mn@doi [\apj]
  {10.3847/1538-4357/acafe7}, \href
  {https://ui.adsabs.harvard.edu/abs/2023ApJ...946...19D} {946, 19}

\bibitem[\protect\citeauthoryear{{Fabian}, {Rees}, {Stella}  \&
  {White}}{{Fabian} et~al.}{1989}]{Fabian1989}
{Fabian} A.~C.,  {Rees} M.~J.,  {Stella} L.,   {White} N.~E.,  1989, \mn@doi
  [\mnras] {10.1093/mnras/238.3.729}, \href
  {https://ui.adsabs.harvard.edu/abs/1989MNRAS.238..729F} {238, 729}

\bibitem[\protect\citeauthoryear{{Farr}, {Sravan}, {Cantrell}, {Kreidberg},
  {Bailyn}, {Mandel}  \& {Kalogera}}{{Farr} et~al.}{2011}]{Farr2011}
{Farr} W.~M.,  {Sravan} N.,  {Cantrell} A.,  {Kreidberg} L.,  {Bailyn} C.~D.,
  {Mandel} I.,   {Kalogera} V.,  2011, \mn@doi [\apj]
  {10.1088/0004-637X/741/2/103}, \href
  {https://ui.adsabs.harvard.edu/abs/2011ApJ...741..103F} {741, 103}

\bibitem[\protect\citeauthoryear{{Fender} \& {Belloni}}{{Fender} \&
  {Belloni}}{2004}]{Fender2004}
{Fender} R.,  {Belloni} T.,  2004, \mn@doi [\araa]
  {10.1146/annurev.astro.42.053102.134031}, \href
  {https://ui.adsabs.harvard.edu/abs/2004ARA&A..42..317F} {42, 317}

\bibitem[\protect\citeauthoryear{{Fender} \& {Belloni}}{{Fender} \&
  {Belloni}}{2012}]{Fender2012}
{Fender} R.,  {Belloni} T.,  2012, \mn@doi [Science] {10.1126/science.1221790},
  \href {https://ui.adsabs.harvard.edu/abs/2012Sci...337..540F} {337, 540}

\bibitem[\protect\citeauthoryear{{Fishbach} \& {Kalogera}}{{Fishbach} \&
  {Kalogera}}{2022}]{Fishbach2022}
{Fishbach} M.,  {Kalogera} V.,  2022, \mn@doi [\apjl]
  {10.3847/2041-8213/ac64a5}, \href
  {https://ui.adsabs.harvard.edu/abs/2022ApJ...929L..26F} {929, L26}

\bibitem[\protect\citeauthoryear{{Fryer} \& {Kalogera}}{{Fryer} \&
  {Kalogera}}{2001}]{Fryer2001}
{Fryer} C.~L.,  {Kalogera} V.,  2001, \mn@doi [\apj] {10.1086/321359}, \href
  {https://ui.adsabs.harvard.edu/abs/2001ApJ...554..548F} {554, 548}

\bibitem[\protect\citeauthoryear{{Fryer}, {Belczynski}, {Wiktorowicz},
  {Dominik}, {Kalogera}  \& {Holz}}{{Fryer} et~al.}{2012}]{Fryer2012}
{Fryer} C.~L.,  {Belczynski} K.,  {Wiktorowicz} G.,  {Dominik} M.,  {Kalogera}
  V.,   {Holz} D.~E.,  2012, \mn@doi [\apj] {10.1088/0004-637X/749/1/91}, \href
  {https://ui.adsabs.harvard.edu/abs/2012ApJ...749...91F} {749, 91}

\bibitem[\protect\citeauthoryear{{F{\"u}rst} et~al.,}{{F{\"u}rst}
  et~al.}{2016}]{Fuerst2016}
{F{\"u}rst} F.,  et~al., 2016, \mn@doi [\apj] {10.3847/0004-637X/828/1/34},
  \href {https://ui.adsabs.harvard.edu/abs/2016ApJ...828...34F} {828, 34}

\bibitem[\protect\citeauthoryear{{Garc{\'\i}a}, {Dauser}, {Reynolds},
  {Kallman}, {McClintock}, {Wilms}  \& {Eikmann}}{{Garc{\'\i}a}
  et~al.}{2013}]{Garcia2013}
{Garc{\'\i}a} J.,  {Dauser} T.,  {Reynolds} C.~S.,  {Kallman} T.~R.,
  {McClintock} J.~E.,  {Wilms} J.,   {Eikmann} W.,  2013, \mn@doi [\apj]
  {10.1088/0004-637X/768/2/146}, \href
  {https://ui.adsabs.harvard.edu/abs/2013ApJ...768..146G} {768, 146}

\bibitem[\protect\citeauthoryear{{Garc{\'\i}a} et~al.,}{{Garc{\'\i}a}
  et~al.}{2014a}]{Garcia2014}
{Garc{\'\i}a} J.,  et~al., 2014a, \mn@doi [\apj] {10.1088/0004-637X/782/2/76},
  \href {https://ui.adsabs.harvard.edu/abs/2014ApJ...782...76G} {782, 76}

\bibitem[\protect\citeauthoryear{{Garc{\'\i}a}, {McClintock}, {Steiner},
  {Remillard}  \& {Grinberg}}{{Garc{\'\i}a} et~al.}{2014b}]{Garcia2014pcacorr}
{Garc{\'\i}a} J.~A.,  {McClintock} J.~E.,  {Steiner} J.~F.,  {Remillard} R.~A.,
    {Grinberg} V.,  2014b, \mn@doi [\apj] {10.1088/0004-637X/794/1/73}, \href
  {https://ui.adsabs.harvard.edu/abs/2014ApJ...794...73G} {794, 73}

\bibitem[\protect\citeauthoryear{{Garc{\'{\i}}a}, {Steiner}, {McClintock},
  {Remillard}, {Grinberg}  \& {Dauser}}{{Garc{\'{\i}}a}
  et~al.}{2015}]{Garcia2015}
{Garc{\'{\i}}a} J.~A.,  {Steiner} J.~F.,  {McClintock} J.~E.,  {Remillard}
  R.~A.,  {Grinberg} V.,   {Dauser} T.,  2015, \mn@doi [\apj]
  {10.1088/0004-637X/813/2/84}, \href
  {http://adsabs.harvard.edu/abs/2015ApJ...813...84G} {813, 84}

\bibitem[\protect\citeauthoryear{{Garc{\'\i}a}, {Fabian}, {Kallman}, {Dauser},
  {Parker}, {McClintock}, {Steiner}  \& {Wilms}}{{Garc{\'\i}a}
  et~al.}{2016}]{Garcia2016}
{Garc{\'\i}a} J.~A.,  {Fabian} A.~C.,  {Kallman} T.~R.,  {Dauser} T.,  {Parker}
  M.~L.,  {McClintock} J.~E.,  {Steiner} J.~F.,   {Wilms} J.,  2016, \mn@doi
  [\mnras] {10.1093/mnras/stw1696}, \href
  {https://ui.adsabs.harvard.edu/abs/2016MNRAS.462..751G} {462, 751}

\bibitem[\protect\citeauthoryear{{Garc{\'\i}a} et~al.,}{{Garc{\'\i}a}
  et~al.}{2019}]{Garcia2019}
{Garc{\'\i}a} J.~A.,  et~al., 2019, \mn@doi [\apj] {10.3847/1538-4357/aaf739},
  \href {https://ui.adsabs.harvard.edu/abs/2019ApJ...871...88G} {871, 88}

\bibitem[\protect\citeauthoryear{{Harrison} et~al.,}{{Harrison}
  et~al.}{2013}]{Harrison2013}
{Harrison} F.~A.,  et~al., 2013, \mn@doi [\apj] {10.1088/0004-637X/770/2/103},
  \href {https://ui.adsabs.harvard.edu/abs/2013ApJ...770..103H} {770, 103}

\bibitem[\protect\citeauthoryear{{Ingram} \& {Motta}}{{Ingram} \&
  {Motta}}{2014}]{Ingram2014}
{Ingram} A.,  {Motta} S.,  2014, \mn@doi [\mnras] {10.1093/mnras/stu1585},
  \href {https://ui.adsabs.harvard.edu/abs/2014MNRAS.444.2065I} {444, 2065}

\bibitem[\protect\citeauthoryear{{Ingram} \& {Motta}}{{Ingram} \&
  {Motta}}{2019}]{Ingram2019review}
{Ingram} A.~R.,  {Motta} S.~E.,  2019, \mn@doi [\nar]
  {10.1016/j.newar.2020.101524}, \href
  {https://ui.adsabs.harvard.edu/abs/2019NewAR..8501524I} {85, 101524}

\bibitem[\protect\citeauthoryear{{Ingram}, {Done}  \& {Fragile}}{{Ingram}
  et~al.}{2009}]{Ingram2009}
{Ingram} A.,  {Done} C.,   {Fragile} P.~C.,  2009, \mn@doi [\mnras]
  {10.1111/j.1745-3933.2009.00693.x}, \href
  {https://ui.adsabs.harvard.edu/abs/2009MNRAS.397L.101I} {397, L101}

\bibitem[\protect\citeauthoryear{{Ingram}, {van der Klis}, {Middleton}, {Done},
  {Altamirano}, {Heil}, {Uttley}  \& {Axelsson}}{{Ingram}
  et~al.}{2016}]{Ingram2016}
{Ingram} A.,  {van der Klis} M.,  {Middleton} M.,  {Done} C.,  {Altamirano} D.,
   {Heil} L.,  {Uttley} P.,   {Axelsson} M.,  2016, \mn@doi [\mnras]
  {10.1093/mnras/stw1245}, \href
  {https://ui.adsabs.harvard.edu/abs/2016MNRAS.461.1967I} {461, 1967}

\bibitem[\protect\citeauthoryear{{Ingram}, {van der Klis}, {Middleton},
  {Altamirano}  \& {Uttley}}{{Ingram} et~al.}{2017}]{Ingram2017h1743}
{Ingram} A.,  {van der Klis} M.,  {Middleton} M.,  {Altamirano} D.,   {Uttley}
  P.,  2017, \mn@doi [\mnras] {10.1093/mnras/stw2581}, \href
  {https://ui.adsabs.harvard.edu/abs/2017MNRAS.464.2979I} {464, 2979}

\bibitem[\protect\citeauthoryear{{Ingram}, {Mastroserio}, {Dauser},
  {Hovenkamp}, {van der Klis}  \& {Garc{\'\i}a}}{{Ingram}
  et~al.}{2019}]{Ingram2019reltrans}
{Ingram} A.,  {Mastroserio} G.,  {Dauser} T.,  {Hovenkamp} P.,  {van der Klis}
  M.,   {Garc{\'\i}a} J.~A.,  2019, \mn@doi [\mnras] {10.1093/mnras/stz1720},
  \href {https://ui.adsabs.harvard.edu/abs/2019MNRAS.488..324I} {488, 324}

\bibitem[\protect\citeauthoryear{{Ingram} et~al.,}{{Ingram}
  et~al.}{2022}]{Ingram2022}
{Ingram} A.,  et~al., 2022, \mn@doi [\mnras] {10.1093/mnras/stab2950}, \href
  {https://ui.adsabs.harvard.edu/abs/2022MNRAS.509..619I} {509, 619}

\bibitem[\protect\citeauthoryear{{Jansen} et~al.,}{{Jansen}
  et~al.}{2001}]{Jansen2001}
{Jansen} F.,  et~al., 2001, \mn@doi [\aap] {10.1051/0004-6361:20000036}, \href
  {https://ui.adsabs.harvard.edu/abs/2001A&A...365L...1J} {365, L1}

\bibitem[\protect\citeauthoryear{{Jiang}, {Fabian}, {Wang}, {Walton},
  {Garc{\'\i}a}, {Parker}, {Steiner}  \& {Tomsick}}{{Jiang}
  et~al.}{2019a}]{Jiang2019a}
{Jiang} J.,  {Fabian} A.~C.,  {Wang} J.,  {Walton} D.~J.,  {Garc{\'\i}a} J.~A.,
   {Parker} M.~L.,  {Steiner} J.~F.,   {Tomsick} J.~A.,  2019a, \mn@doi
  [\mnras] {10.1093/mnras/stz095}, \href
  {https://ui.adsabs.harvard.edu/abs/2019MNRAS.484.1972J} {484, 1972}

\bibitem[\protect\citeauthoryear{{Jiang} et~al.,}{{Jiang}
  et~al.}{2019b}]{Jiang2019b}
{Jiang} J.,  et~al., 2019b, \mn@doi [\mnras] {10.1093/mnras/stz2326}, \href
  {https://ui.adsabs.harvard.edu/abs/2019MNRAS.489.3436J} {489, 3436}

\bibitem[\protect\citeauthoryear{{Jonker}, {Kaur}, {Stone}  \&
  {Torres}}{{Jonker} et~al.}{2021}]{Jonker2021}
{Jonker} P.~G.,  {Kaur} K.,  {Stone} N.,   {Torres} M. A.~P.,  2021, \mn@doi
  [\apj] {10.3847/1538-4357/ac2839}, \href
  {https://ui.adsabs.harvard.edu/abs/2021ApJ...921..131J} {921, 131}

\bibitem[\protect\citeauthoryear{{Kammoun} et~al.,}{{Kammoun}
  et~al.}{2024}]{Kammoun2024}
{Kammoun} E.,  et~al., 2024, \mn@doi [Frontiers in Astronomy and Space
  Sciences] {10.3389/fspas.2023.1308056}, \href
  {https://ui.adsabs.harvard.edu/abs/2024FrASS..1008056K} {10, 1308056}

\bibitem[\protect\citeauthoryear{{Kolehmainen}, {Done}  \& {D{\'\i}az
  Trigo}}{{Kolehmainen} et~al.}{2014}]{Kolehmainen2014}
{Kolehmainen} M.,  {Done} C.,   {D{\'\i}az Trigo} M.,  2014, \mn@doi [\mnras]
  {10.1093/mnras/stt1886}, \href
  {https://ui.adsabs.harvard.edu/abs/2014MNRAS.437..316K} {437, 316}

\bibitem[\protect\citeauthoryear{{Krawczynski} et~al.,}{{Krawczynski}
  et~al.}{2022}]{Krawczynski2022}
{Krawczynski} H.,  et~al., 2022, \mn@doi [Science] {10.1126/science.add5399},
  \href {https://ui.adsabs.harvard.edu/abs/2022Sci...378..650K} {378, 650}

\bibitem[\protect\citeauthoryear{{Li}, {Zimmerman}, {Narayan}  \&
  {McClintock}}{{Li} et~al.}{2005}]{Li2005}
{Li} L.-X.,  {Zimmerman} E.~R.,  {Narayan} R.,   {McClintock} J.~E.,  2005,
  \mn@doi [\apjs] {10.1086/428089}, \href
  {https://ui.adsabs.harvard.edu/abs/2005ApJS..157..335L} {157, 335}

\bibitem[\protect\citeauthoryear{{Liu} et~al.,}{{Liu} et~al.}{2023}]{Liu2023}
{Liu} H.,  et~al., 2023, \mn@doi [arXiv e-prints] {10.48550/arXiv.2303.10593},
  \href {https://ui.adsabs.harvard.edu/abs/2023arXiv230310593L} {p.
  arXiv:2303.10593}

\bibitem[\protect\citeauthoryear{{Lubow}, {Ogilvie}  \& {Pringle}}{{Lubow}
  et~al.}{2002}]{Lubow2002}
{Lubow} S.~H.,  {Ogilvie} G.~I.,   {Pringle} J.~E.,  2002, \mn@doi [\mnras]
  {10.1046/j.1365-8711.2002.05949.x}, \href
  {https://ui.adsabs.harvard.edu/abs/2002MNRAS.337..706L} {337, 706}

\bibitem[\protect\citeauthoryear{{Madsen}, {Forster}, {Grefenstette},
  {Harrison}  \& {Stern}}{{Madsen} et~al.}{2017}]{Madsen2017}
{Madsen} K.~K.,  {Forster} K.,  {Grefenstette} B.~W.,  {Harrison} F.~A.,
  {Stern} D.,  2017, \mn@doi [\apj] {10.3847/1538-4357/aa6970}, \href
  {https://ui.adsabs.harvard.edu/abs/2017ApJ...841...56M} {841, 56}

\bibitem[\protect\citeauthoryear{{Mastroserio}, {Ingram}  \& {van der
  Klis}}{{Mastroserio} et~al.}{2019}]{Mastroserio2019}
{Mastroserio} G.,  {Ingram} A.,   {van der Klis} M.,  2019, \mn@doi [\mnras]
  {10.1093/mnras/stz1727}, \href
  {https://ui.adsabs.harvard.edu/abs/2019MNRAS.488..348M} {488, 348}

\bibitem[\protect\citeauthoryear{{Matt}, {Perola}  \& {Piro}}{{Matt}
  et~al.}{1991}]{Matt1991}
{Matt} G.,  {Perola} G.~C.,   {Piro} L.,  1991, \aap, \href
  {https://ui.adsabs.harvard.edu/abs/1991A&A...247...25M} {247, 25}

\bibitem[\protect\citeauthoryear{{McClintock}, {Shafee}, {Narayan},
  {Remillard}, {Davis}  \& {Li}}{{McClintock} et~al.}{2006}]{McClintock2006}
{McClintock} J.~E.,  {Shafee} R.,  {Narayan} R.,  {Remillard} R.~A.,  {Davis}
  S.~W.,   {Li} L.-X.,  2006, \mn@doi [\apj] {10.1086/508457}, \href
  {https://ui.adsabs.harvard.edu/abs/2006ApJ...652..518M} {652, 518}

\bibitem[\protect\citeauthoryear{{McClintock}, {Narayan}  \&
  {Steiner}}{{McClintock} et~al.}{2014}]{McClintock2014}
{McClintock} J.~E.,  {Narayan} R.,   {Steiner} J.~F.,  2014, \mn@doi [\ssr]
  {10.1007/s11214-013-0003-9}, \href
  {https://ui.adsabs.harvard.edu/abs/2014SSRv..183..295M} {183, 295}

\bibitem[\protect\citeauthoryear{{Miller-Jones} et~al.,}{{Miller-Jones}
  et~al.}{2021}]{MillerJones2021}
{Miller-Jones} J. C.~A.,  et~al., 2021, \mn@doi [Science]
  {10.1126/science.abb3363}, \href
  {https://ui.adsabs.harvard.edu/abs/2021Sci...371.1046M} {371, 1046}

\bibitem[\protect\citeauthoryear{{Molla}, {Chakrabarti}, {Debnath}  \&
  {Mondal}}{{Molla} et~al.}{2017}]{Molla2017}
{Molla} A.~A.,  {Chakrabarti} S.~K.,  {Debnath} D.,   {Mondal} S.,  2017,
  \mn@doi [\apj] {10.3847/1538-4357/834/1/88}, \href
  {https://ui.adsabs.harvard.edu/abs/2017ApJ...834...88M} {834, 88}

\bibitem[\protect\citeauthoryear{{Mondal}}{{Mondal}}{2010}]{Mondal2010}
{Mondal} S.,  2010, \mn@doi [\apj] {10.1088/0004-637X/708/2/1442}, \href
  {https://ui.adsabs.harvard.edu/abs/2010ApJ...708.1442M} {708, 1442}

\bibitem[\protect\citeauthoryear{{Nathan} et~al.,}{{Nathan}
  et~al.}{2022}]{Nathan2022grs1915}
{Nathan} E.,  et~al., 2022, \mn@doi [\mnras] {10.1093/mnras/stab3803}, \href
  {https://ui.adsabs.harvard.edu/abs/2022MNRAS.511..255N} {511, 255}

\bibitem[\protect\citeauthoryear{{Novikov} \& {Thorne}}{{Novikov} \&
  {Thorne}}{1973}]{Novikov1973}
{Novikov} I.~D.,  {Thorne} K.~S.,  1973, in Black Holes (Les Astres Occlus). pp
  343--450

\bibitem[\protect\citeauthoryear{{Orosz}, {McClintock}, {Aufdenberg},
  {Remillard}, {Reid}, {Narayan}  \& {Gou}}{{Orosz} et~al.}{2011}]{Orosz2011}
{Orosz} J.~A.,  {McClintock} J.~E.,  {Aufdenberg} J.~P.,  {Remillard} R.~A.,
  {Reid} M.~J.,  {Narayan} R.,   {Gou} L.,  2011, \mn@doi [\apj]
  {10.1088/0004-637X/742/2/84}, \href
  {https://ui.adsabs.harvard.edu/abs/2011ApJ...742...84O} {742, 84}

\bibitem[\protect\citeauthoryear{{{\"O}zel}, {Psaltis}, {Narayan}  \&
  {McClintock}}{{{\"O}zel} et~al.}{2010}]{Ozel2010}
{{\"O}zel} F.,  {Psaltis} D.,  {Narayan} R.,   {McClintock} J.~E.,  2010,
  \mn@doi [\apj] {10.1088/0004-637X/725/2/1918}, \href
  {https://ui.adsabs.harvard.edu/abs/2010ApJ...725.1918O} {725, 1918}

\bibitem[\protect\citeauthoryear{{Parker} et~al.,}{{Parker}
  et~al.}{2016}]{Parker2016}
{Parker} M.~L.,  et~al., 2016, \mn@doi [\apjl] {10.3847/2041-8205/821/1/L6},
  \href {https://ui.adsabs.harvard.edu/abs/2016ApJ...821L...6P} {821, L6}

\bibitem[\protect\citeauthoryear{{P{\'e}tri}}{{P{\'e}tri}}{2008}]{Petri2008}
{P{\'e}tri} J.,  2008, \mn@doi [\apss] {10.1007/s10509-008-9916-2}, \href
  {https://ui.adsabs.harvard.edu/abs/2008Ap&SS.318..181P} {318, 181}

\bibitem[\protect\citeauthoryear{{Poutanen}, {Veledina}  \&
  {Zdziarski}}{{Poutanen} et~al.}{2018}]{Poutanen2018}
{Poutanen} J.,  {Veledina} A.,   {Zdziarski} A.~A.,  2018, \mn@doi [\aap]
  {10.1051/0004-6361/201732345}, \href
  {https://ui.adsabs.harvard.edu/abs/2018A&A...614A..79P} {614, A79}

\bibitem[\protect\citeauthoryear{{Remillard} \& {McClintock}}{{Remillard} \&
  {McClintock}}{2006}]{Remillard2006}
{Remillard} R.~A.,  {McClintock} J.~E.,  2006, \mn@doi [\araa]
  {10.1146/annurev.astro.44.051905.092532}, \href
  {https://ui.adsabs.harvard.edu/abs/2006ARA&A..44...49R} {44, 49}

\bibitem[\protect\citeauthoryear{Ross \& Fabian}{Ross \&
  Fabian}{2005}]{Ross2005}
Ross R.,  Fabian A.,  2005, Monthly Notices of the Royal Astronomical Society,
  358, 211

\bibitem[\protect\citeauthoryear{{Shakura} \& {Sunyaev}}{{Shakura} \&
  {Sunyaev}}{1973}]{Shakura1973}
{Shakura} N.~I.,  {Sunyaev} R.~A.,  1973, in {Bradt} H.,  {Giacconi} R.,  eds,
  IAU Symposium Vol. 55, X- and Gamma-Ray Astronomy. p.~155

\bibitem[\protect\citeauthoryear{{Shaposhnikov} \& {Titarchuk}}{{Shaposhnikov}
  \& {Titarchuk}}{2009}]{Shaposhnikov2009}
{Shaposhnikov} N.,  {Titarchuk} L.,  2009, \mn@doi [\apj]
  {10.1088/0004-637X/699/1/453}, \href
  {https://ui.adsabs.harvard.edu/abs/2009ApJ...699..453S} {699, 453}

\bibitem[\protect\citeauthoryear{{Shreeram} \& {Ingram}}{{Shreeram} \&
  {Ingram}}{2020}]{Shreeram2020}
{Shreeram} S.,  {Ingram} A.,  2020, \mn@doi [\mnras] {10.1093/mnras/stz3455},
  \href {https://ui.adsabs.harvard.edu/abs/2020MNRAS.492..405S} {492, 405}

\bibitem[\protect\citeauthoryear{{Sreehari}, {Iyer}, {Radhika}, {Nandi}  \&
  {Mandal}}{{Sreehari} et~al.}{2019}]{Sreehari2019}
{Sreehari} H.,  {Iyer} N.,  {Radhika} D.,  {Nandi} A.,   {Mandal} S.,  2019,
  \mn@doi [Advances in Space Research] {10.1016/j.asr.2018.10.042}, \href
  {https://ui.adsabs.harvard.edu/abs/2019AdSpR..63.1374S} {63, 1374}

\bibitem[\protect\citeauthoryear{{Steiner}, {McClintock}, {Remillard},
  {Narayan}  \& {Gou}}{{Steiner} et~al.}{2009}]{Steiner2009}
{Steiner} J.~F.,  {McClintock} J.~E.,  {Remillard} R.~A.,  {Narayan} R.,
  {Gou} L.,  2009, \mn@doi [\apjl] {10.1088/0004-637X/701/2/L83}, \href
  {https://ui.adsabs.harvard.edu/abs/2009ApJ...701L..83S} {701, L83}

\bibitem[\protect\citeauthoryear{{Steiner}, {McClintock}  \& {Reid}}{{Steiner}
  et~al.}{2012}]{Steiner2012}
{Steiner} J.~F.,  {McClintock} J.~E.,   {Reid} M.~J.,  2012, \mn@doi [\apjl]
  {10.1088/2041-8205/745/1/L7}, \href
  {https://ui.adsabs.harvard.edu/abs/2012ApJ...745L...7S} {745, L7}

\bibitem[\protect\citeauthoryear{{Sunyaev} \& {Truemper}}{{Sunyaev} \&
  {Truemper}}{1979}]{Sunyaev1979}
{Sunyaev} R.~A.,  {Truemper} J.,  1979, \mn@doi [\nat] {10.1038/279506a0},
  \href {https://ui.adsabs.harvard.edu/abs/1979Natur.279..506S} {279, 506}

\bibitem[\protect\citeauthoryear{{Tetarenko}, {Sivakoff}, {Heinke}  \&
  {Gladstone}}{{Tetarenko} et~al.}{2016}]{Tetarenko2016}
{Tetarenko} B.~E.,  {Sivakoff} G.~R.,  {Heinke} C.~O.,   {Gladstone} J.~C.,
  2016, \mn@doi [\apjs] {10.3847/0067-0049/222/2/15}, \href
  {https://ui.adsabs.harvard.edu/abs/2016ApJS..222...15T} {222, 15}

\bibitem[\protect\citeauthoryear{{Thorne} \& {Price}}{{Thorne} \&
  {Price}}{1975}]{Thorne1975}
{Thorne} K.~S.,  {Price} R.~H.,  1975, \mn@doi [\apjl] {10.1086/181720}, \href
  {https://ui.adsabs.harvard.edu/abs/1975ApJ...195L.101T} {195, L101}

\bibitem[\protect\citeauthoryear{{Tomsick} et~al.,}{{Tomsick}
  et~al.}{2018}]{Tomsick2018}
{Tomsick} J.~A.,  et~al., 2018, \mn@doi [\apj] {10.3847/1538-4357/aaaab1},
  \href {https://ui.adsabs.harvard.edu/abs/2018ApJ...855....3T} {855, 3}

\bibitem[\protect\citeauthoryear{{Tursunov} \& {Kolo{\v{s}}}}{{Tursunov} \&
  {Kolo{\v{s}}}}{2018}]{Tursunov2018}
{Tursunov} A.~A.,  {Kolo{\v{s}}} M.,  2018, \mn@doi [Physics of Atomic Nuclei]
  {10.1134/S1063778818020187}, \href
  {https://ui.adsabs.harvard.edu/abs/2018PAN....81..279T} {81, 279}

\bibitem[\protect\citeauthoryear{{Verbunt}, {Igoshev}  \& {Cator}}{{Verbunt}
  et~al.}{2017}]{Verbunt2017}
{Verbunt} F.,  {Igoshev} A.,   {Cator} E.,  2017, \mn@doi [\aap]
  {10.1051/0004-6361/201731518}, \href
  {https://ui.adsabs.harvard.edu/abs/2017A&A...608A..57V} {608, A57}

\bibitem[\protect\citeauthoryear{{Wilms}, {Allen}  \& {McCray}}{{Wilms}
  et~al.}{2000}]{Wilms2000}
{Wilms} J.,  {Allen} A.,   {McCray} R.,  2000, \mn@doi [\apj] {10.1086/317016},
  \href {https://ui.adsabs.harvard.edu/abs/2000ApJ...542..914W} {542, 914}

\bibitem[\protect\citeauthoryear{Yang \& Wang}{Yang \& Wang}{2013}]{Yang2013}
Yang X.,  Wang J.,  2013, The Astrophysical Journal Supplement Series, 207, 6

\bibitem[\protect\citeauthoryear{{Zdziarski} \& {De Marco}}{{Zdziarski} \& {De
  Marco}}{2020}]{Zdziarski2020}
{Zdziarski} A.~A.,  {De Marco} B.,  2020, \mn@doi [\apjl]
  {10.3847/2041-8213/ab9899}, \href
  {https://ui.adsabs.harvard.edu/abs/2020ApJ...896L..36Z} {896, L36}

\bibitem[\protect\citeauthoryear{{Zdziarski}, {Johnson}  \&
  {Magdziarz}}{{Zdziarski} et~al.}{1996}]{Zdziarski1996}
{Zdziarski} A.~A.,  {Johnson} W.~N.,   {Magdziarz} P.,  1996, \mnras, \href
  {http://adsabs.harvard.edu/abs/1996MNRAS.283..193Z} {283, 193}

\bibitem[\protect\citeauthoryear{{Zdziarski}, {You}, {Szanecki}, {Li}  \&
  {Ge}}{{Zdziarski} et~al.}{2022}]{Zdziarski2022}
{Zdziarski} A.~A.,  {You} B.,  {Szanecki} M.,  {Li} X.-B.,   {Ge} M.,  2022,
  \mn@doi [\apj] {10.3847/1538-4357/ac54a7}, \href
  {https://ui.adsabs.harvard.edu/abs/2022ApJ...928...11Z} {928, 11}

\bibitem[\protect\citeauthoryear{{Zhang}, {Cui}  \& {Chen}}{{Zhang}
  et~al.}{1997}]{Zhang1997}
{Zhang} S.~N.,  {Cui} W.,   {Chen} W.,  1997, \mn@doi [\apjl] {10.1086/310705},
  \href {https://ui.adsabs.harvard.edu/abs/1997ApJ...482L.155Z} {482, L155}

\bibitem[\protect\citeauthoryear{{van den Eijnden}, {Bagnoli}, {Degenaar},
  {Lohfink}, {Parker}, {in 't Zand}  \& {Fabian}}{{van den Eijnden}
  et~al.}{2017}]{vandenEijnden2017}
{van den Eijnden} J.,  {Bagnoli} T.,  {Degenaar} N.,  {Lohfink} A.~M.,
  {Parker} M.~L.,  {in 't Zand} J.~J.~M.,   {Fabian} A.~C.,  2017, \mn@doi
  [\mnras] {10.1093/mnrasl/slw244}, \href
  {https://ui.adsabs.harvard.edu/abs/2017MNRAS.466L..98V} {466, L98}

\makeatother
\end{thebibliography}

%%%%%%%%%%%%%%%%%%%%%%%%%%%%%%%%%%%%%%%%%%%%%%%%%%

%%%%%%%%%%%%%%%%% APPENDICES %%%%%%%%%%%%%%%%%%%%%

%\appendix

%\section{Some extra material}

%If you want to present additional material which would interrupt the flow of the main paper, it can be placed in an Appendix which appears after the list of references.

%%%%%%%%%%%%%%%%%%%%%%%%%%%%%%%%%%%%%%%%%%%%%%%%%%

% Don't change these lines
\bsp	% typesetting comment
\label{lastpage}
\end{document}